\documentclass[10pt,preprint]{aastex}

\begin{document}

\shortauthors{Welty}
\shorttitle{Variable IS Absorption}


\title{Monitoring the Variable Interstellar Absorption toward HD~219188 with {\it HST}/STIS\altaffilmark{1}}

\author{Daniel E. Welty}
\affil{University of Chicago, Astronomy and Astrophysics Center, 5640 S. Ellis Ave., Chicago, IL  60637}
\email{welty@oddjob.uchicago.edu}

\altaffiltext{1}{Based on observations made with the NASA/ESA {\it Hubble Space Telescope}, which is operated by the Association of Universities for Research in Astronomy, Inc., under NASA contract NAS5-26555.
Based in part on observations obtained with the Apache Point Observatory 3.5m telescope, which is owned and operated by the Astrophysical Research Consortium.}

\begin{abstract}

We discuss the results of continued optical and UV spectroscopic monitoring of the variable intermediate-velocity (IV) absorption at $v_{\odot}$ = $-$38 km~s$^{-1}$ toward the low halo star HD~219188. 
After reaching maxima in mid-2000, the column densities of both \ion{Na}{1} and \ion{Ca}{2} in that IV component declined by factors $\ga$ 2 by the end of 2006. 
Comparisons between {\it HST}/STIS echelle spectra obtained in 2001, 2003, and 2004 and {\it HST}/GHRS echelle spectra obtained in 1994--1995 indicate the following:
(1) The absorption from the dominant species \ion{S}{2}, \ion{O}{1}, \ion{Si}{2}, and \ion{Fe}{2} is roughly constant in all four sets of spectra --- suggesting that the total $N$(H) ($\sim$ 6~$\times$~10$^{17}$~cm$^{-2}$) and the (mild) depletions have not changed significantly over a period of nearly ten years.
(2) The column densities of the trace species \ion{C}{1} (both ground and excited fine-structure states) and of the excited state \ion{C}{2}* all increased by factors of 2--5 between 1995 and 2001 --- implying increases in the local hydrogen density $n_{\rm H}$ (from about 20 cm$^{-3}$ to about 45 cm$^{-3}$, for $T$~=~100~K) and in the electron density $n_e$ (by a factor $\ga$ 3) over that 6-year period.
(3) The column densities of \ion{C}{1} and \ion{C}{2}* --- and the corresponding inferred $n_{\rm H}$ and $n_e$ --- then decreased slightly between 2001 and 2004.
(4) The changes in \ion{C}{1} and \ion{C}{2}* are very similar to those seen for \ion{Na}{1} and \ion{Ca}{2}.
The relatively low total $N$(H) and the modest $n_{\rm H}$ suggest that the $-$38 km~s$^{-1}$ cloud toward HD~219188 is not a very dense knot or filament.  
Partial ionization of hydrogen appears to be responsible for the enhanced abundances of \ion{Na}{1}, \ion{C}{1}, \ion{Ca}{2}, and \ion{C}{2}*.
In this case, the variations in those species appear to reflect differences in density and ionization [and not $N$(H)] over scales of tens of AU.

\end{abstract}


\section{Introduction}
\label{sec-intro}

Temporal variations in the strengths and/or velocities of interstellar absorption lines have now been observed in a number of sight lines sampling different regions in the Galactic ISM; see, e.g., the reviews by Crawford (2003) and Lauroesch (2007).
The changes in the profiles of the \ion{Na}{1} and/or \ion{Ca}{2} lines observed toward several stars in the direction of the Vela supernova remnant (Hobbs et al. 1991; Danks \& Sembach 1995; Cha \& Sembach 2000) are perhaps not surprising, given the presence of gas at moderately high velocities in that energetic environment.
For some of those variable features, high $N$(\ion{Ca}{2})/$N$(\ion{Na}{1}) ratios suggest significant shock processing of the interstellar dust grains.
In a number of other sight lines, however, the narrow \ion{Na}{1} line widths ($b$ $\sim$ 0.3--0.6 km s$^{-1}$, or FWHM $\sim$ 0.5--1.0 km~s$^{-1}$) and/or low $N$(\ion{Ca}{2})/$N$(\ion{Na}{1}) ratios characterizing the variable components suggest that the gas is relatively cool and quiescent (Blades et al. 1997; Price et al. 2000; Crawford et al. 2000; Lauroesch et al. 2000; Welty \& Fitzpatrick 2001; Lauroesch \& Meyer 2003).
Variations in CH in diffuse molecular gas have been observed toward AE Aur (HD 34078; Rollinde et al. 2003); corresponding changes in H$_2$ have not been seen, however (Boiss\'{e} et al. 2005).
In at least some of those cases, the variations appear to be due to significant transverse motion of the target stars, such that the lines of sight sample different paths through the intervening (inhomogeneous) interstellar clouds at different times.
The observed temporal variations would then imply corresponding spatial variations in cloud properties over scales typically of tens of AU.

Variations in \ion{Na}{1} absorption over somewhat larger spatial scales (10$^2$--10$^4$ AU) have been observed toward a number of binary/multiple star systems (Watson \& Meyer 1996) and toward both open and globular clusters (Langer et al. 1990; Kemp et al. 1993; Meyer \& Lauroesch 1999; Points et al. 2004).
Similar variations in \ion{K}{1}, CH, and/or CN (which trace somewhat denser and/or higher column density gas) have also been noted in a few cases (Lauroesch \& Meyer 1999; Pan et al. 2001).
The seeming ubiquity of such variations has suggested that sub-parsec-scale structure is common in the diffuse Galactic ISM.  
Because \ion{Na}{1} and \ion{K}{1} generally are trace ionization stages, however, it has not been clear whether that structure is in the overall hydrogen column density or just in the local physical conditions affecting the ionization balance.
While small-scale spatial and/or temporal variations in $N$(\ion{H}{1}) have been inferred from VLBI observations of extended extragalactic radio sources (e.g., Dieter et al. 1976; Diamond et al. 1989; Faison et al. 1998; Brogan et al. 2005) and from multi-epoch observations of pulsars showing significant proper motions (Frail et al. 1994), a recent re-examination of some of the pulsar sight lines has suggested that \ion{H}{1} variations on scales less than about 100 AU might not be common (Stanimirovi\'{c} et al. 2003; see also Johnston et al. 2003).

This apparent small-scale structure in predominantly neutral interstellar gas has been difficult to understand. 
If the differences in \ion{Na}{1} are due to variations in total hydrogen column density, with ``typical'' $N$(\ion{Na}{1})/$N$(H) ratios, the implied local hydrogen densities $n_{\rm H}$ would generally be of order several thousand per cm$^3$.
Similar values have been inferred from the observations of small-scale variations in \ion{H}{1} toward both pulsars and extragalactic radio sources.
Such high densities are difficult to reconcile with clouds in thermal pressure equilibrium at typical interstellar values of $P/k$ = $n_{\rm H}T$ $\sim$ 2500 cm$^{-3}$K (Jenkins \& Tripp 2001).
Several solutions for this ``density problem'' have been proposed:
(1) The differences may reflect a population of cold, dense filaments or sheets, embedded in warmer, less dense neutral gas and containing 10--30\% of the total column density of cold, neutral gas (Heiles 1997).
(2) The differences may just be due to statistical fluctuations in the distribution of material at all scales along the sight lines (Deshpande 2003).
(3) The differences in the trace ions might be due to small-scale variations in density and/or ionization (e.g., Lauroesch et al. 1998; Welty \& Fitzpatrick 2001) --- though variations in ionization would not account for the differences seen in \ion{H}{1}.
Unfortunately, high-resolution UV spectra generally have not been available for sight lines exhibiting variability in \ion{Na}{1} (or \ion{H}{1}) --- so there has been little direct information on the physical conditions in the variable components.

The line of sight toward \object{HD~219188}, a B0.5 Ib-II(n) star located about 2800 pc from the Sun at $(l,b)$ $\sim$ (83\arcdeg, $-$50\arcdeg) (Albert 1983), has displayed some of the most striking temporal variations in \ion{Na}{1} absorption yet observed (Welty \& Fitzpatrick 2001).  
Sometime between 1980.75 and 1997.77, a fairly strong, narrow \ion{Na}{1} component appeared at $v_{\sun}$ $\sim$ $-$38 km~s$^{-1}$ toward HD~219188, then continued to strengthen, by an additional factor of 2--3, between 1997.77 and 2000.46.
The line of sight appears to be moving into or through a relatively cold, quiescent intermediate velocity (IV) cloud, as a result of the 13 mas/yr proper motion of HD~219188 (ESA 1997). 
The IV cloud may be associated with an apparent (partial) shell of gas visible in 21 cm emission at velocities between about $-$30 and $-$60 km~s$^{-1}$ (e.g., Hartmann \& Burton 1997).
The variations in \ion{Na}{1} probe transverse length scales of 2--38 AU per year, depending on the (poorly constrained) distance to the IV gas.
The narrow \ion{Na}{1} line width ($b$ $\sim$ 0.55--0.60 km~s$^{-1}$) for the $-$38 km~s$^{-1}$ component implies that the temperature must be less than 490 K.
Analysis of UV spectra obtained with the {\it Hubble Space Telescope} Goddard High-Resolution Spectrograph ({\it HST}/GHRS) in 1994--1995 suggested $N$(H) $\sim$ 5 $\times$ 10$^{17}$ cm$^{-2}$, ``halo cloud'' depletions of silicon and iron, $n_{\rm H}$ $\sim$ 20 cm$^{-3}$, and $n_e$ $\sim$ 0.7--5.9 cm$^{-3}$ (if $T$ $\sim$ 100 K) for the portion of the cloud sampled at that time (Welty \& Fitzpatrick 2001).\footnotemark\
The relatively high fractional ionization, $n_e$/$n_{\rm H}$ $\ga$ 0.035, implies that hydrogen must be partially ionized. 
In this case, the $N$(\ion{Na}{1})/$N$(H) ratio is much higher than ``usual'' --- so that the variations in \ion{Na}{1} do not imply large local pressures or densities.
\footnotetext{The specific numerical values given here are based on a re-analysis of the GHRS data; they differ slightly from those quoted by Welty \& Fitzpatrick (2001).  In this paper, $N$(H) includes \ion{H}{1}, \ion{H}{2}, and H$_2$.}

We have obtained additional optical and UV spectra of HD~219188, at a number of epochs between 2000.80 and 2006.93, in order to monitor the continuing variations in both the column densities of various species and the physical properties of the IV gas in that sight line.
Together with the spectra reported by Welty \& Fitzpatrick (2001), these new spectra provide the most extensive and detailed temporal and spectral coverage reported (to this point) for any sight line with known variations in interstellar absorption.
Section~\ref{sec-obs} describes the newly obtained optical and UV spectra and our analyses of those data.
Section~\ref{sec-res} compares the column densities and physical conditions inferred from the new spectra with previously obtained values.
Section~\ref{sec-disc} discusses some implications of the observed temporal variations in the column densities and physical properties.
Section~\ref{sec-summ} provides a summary of our results.

\section{Observations and Data Analysis}
\label{sec-obs}

\subsection{Optical Spectra}
\label{sec-opt}

Optical spectra of HD~219188 covering the strongest absorption lines from \ion{Na}{1} (5889 and 5895 \AA) and/or \ion{Ca}{2} (3933 and 3968 \AA) have been obtained between 1995.80 and 2006.93 during various observing runs at several different observatories (Table~\ref{tab:datopt}).
Most of the optical spectra discussed by Welty \& Fitzpatrick (2001) were acquired with the Kitt Peak National Observatory 0.9m coud\'{e} feed telescope and echelle spectrograph between 1995.80 and 2000.46, using camera 6 to achieve resolutions of 1.3--1.5 km~s$^{-1}$ (FWHM) (e.g., Welty et al. 1996; Welty \& Hobbs 2001).
Additional spectra, at somewhat lower resolution ($\sim$ 4 km~s$^{-1}$) but higher signal-to-noise (S/N) ratio, were subsequently obtained with the coud\'{e} feed and camera 5 in 2001.86 and 2002.87 by J. Lauroesch.
Several spectra were obtained at the European Southern Observatory (1999.98, 2000.80), using the 3.6m telescope, the coud\'{e} echelle spectrograph (CES), and the Very Long Camera to achieve resolutions of 1.2--1.3 km~s$^{-1}$ (\ion{Na}{1}) and 2.0 km~s$^{-1}$ (\ion{Ca}{2}) (D. E. Welty \& P. A. Crowther, in preparation).
A number of medium resolution (FWHM $\sim$ 8 km s$^{-1}$) but high S/N ratio ($\sim$ 280--630 per resolution element) optical spectra, covering the range from about 3500 to 10000 \AA, were acquired with the Apache Point Observatory 3.5m telescope and the Astrophysical Research Corporation echelle spectrograph (ARCES) between 2002.79 and 2006.93, as part of an extensive observing program designed to investigate the properties of the diffuse interstellar bands (Thorburn et al. 2003; D. G. York et al., in preparation). 

\begin{figure}
\epsscale{0.9}
\plottwo{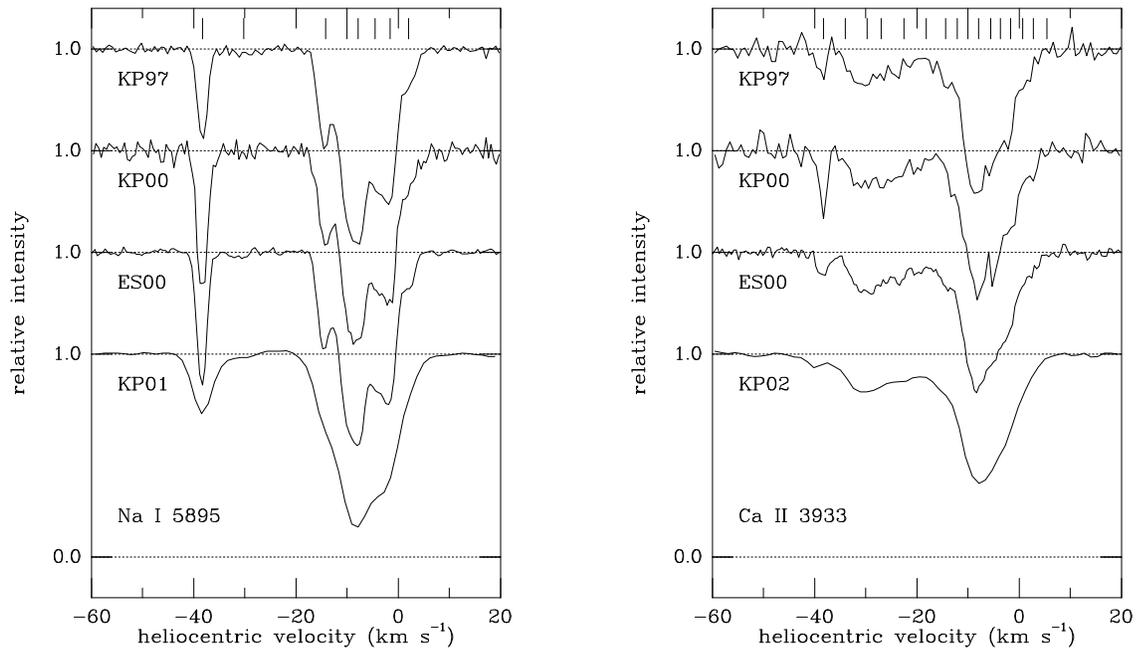}{f1b.eps}
\caption{Selected Na~I ({\it left}) and Ca~II ({\it right}) spectra toward HD 219188, observed with the KPNO coud\'{e} feed and the ESO CES at resolutions of 1.3--4.0 km s$^{-1}$ (see Table~1).
The source and year of each spectrum are indicated; the tick marks above the spectra indicate the components discerned in the fits to the profiles.
Note the clear variations in the strength of the $-$38 km~s$^{-1}$ component in both Na~I and Ca~II, compared to the essentially constant absorption at other velocities.}
\label{fig:naca}
\end{figure}

Standard routines within IRAF were used to remove bias, flat field the spectral order(s) of interest, and extract one-dimensional wavelength-calibrated spectra from the two-dimensional spectral images obtained at KPNO and ESO; locally developed programs then were used to normalize the spectra and to combine the multiple spectra obtained during several of the observing runs.
The spectral resolutions were determined in each case from the widths of the narrow thorium lines [with intrinsic (thermally broadened) FWHM typically about 0.55 km~s$^{-1}$ (Welty et al. 1994)] in the Th-Ar lamp spectra used for wavelength calibration.
The procedures used to extract calibrated one-dimensional spectra from the two-dimensional ARCES data have been described by Thorburn et al. (2003). 
For all the spectra, telluric absorption features near the interstellar \ion{Na}{1} lines were removed using high S/N ratio spectra of bright, nearby stars showing negligible interstellar \ion{Na}{1} absorption.
Several of the higher resolution and/or higher S/N ratio \ion{Na}{1} $\lambda$5895 and \ion{Ca}{2} $\lambda$3933 profiles from KPNO and ESO are shown in Figure~\ref{fig:naca}.
For both species, there are clear variations in the strength of the absorption near $-$38 km~s$^{-1}$ (after accounting for differences in resolution). 

\subsection{UV Spectra}
\label{sec-uv}

High-resolution ultraviolet spectra of HD~219188 have now been obtained with {\it HST} at five epochs between 1994.4 and 2004.5 (Table~\ref{tab:datuv}).
The {\it HST}/GHRS echelle spectra (1994.43 and 1995.37) have been described by Welty \& Fitzpatrick (2001).
New UV spectra were subsequently acquired at three epochs (2001.74, 2003.43, and 2004.42) using the Space Telescope Imaging Spectrograph (STIS) E140H and E230H echelle modes. 
Three wavelength settings, with central wavelengths 1271, 2363, and 2812 \AA\ and covering total ranges of 201--284 \AA, were observed at each epoch.
The STIS data were automatically reduced with the latest version of the CALSTIS pipeline, which includes corrections for scattered light contamination using the procedure described by Valenti et al. (2002). 
Slight (1--2\%) residual flux remaining in the cores of several clearly saturated lines (e.g., of \ion{S}{2} and \ion{Si}{2}) was removed from the spectral segments containing those lines.
Because HD~219188 is relatively bright, all the STIS spectra were obtained through the narrowest (0.10$\times$0.03 arcsec) slit --- which in principle can yield a resolution of about 1.5 km~s$^{-1}$ (Jenkins \& Tripp 2001).
We have chosen, however, to combine adjacent pixels (as in the default CALSTIS processing) in order to increase the S/N ratio in the spectra.
The effective resolution for the STIS spectra was estimated via fits to the narrow, relatively isolated $-$38 km~s$^{-1}$ component for several of the lines of \ion{C}{1}, using a $b$-value for \ion{C}{1} of 0.65 km~s$^{-1}$ (see below).
The resulting instrumental FWHM, about 2.3 km~s$^{-1}$, yields a good fit to the $-$38 km~s$^{-1}$ component for the $\lambda$2852 line of \ion{Mg}{1} as well, using $b$(\ion{Mg}{1}) = $b$(\ion{Na}{1}) = 0.58 km~s$^{-1}$. 
Sections of the spectra near the various interstellar lines of interest (Table~\ref{tab:uvlines}) were normalized using low-order polynomial fits to the adjacent continuum regions. 
The S/N ratios, estimated from the rms fluctuations in those continuum regions, range from about 30 to 70 per resolution element in the individual spectra. 

\begin{figure}
\epsscale{0.9}
\plottwo{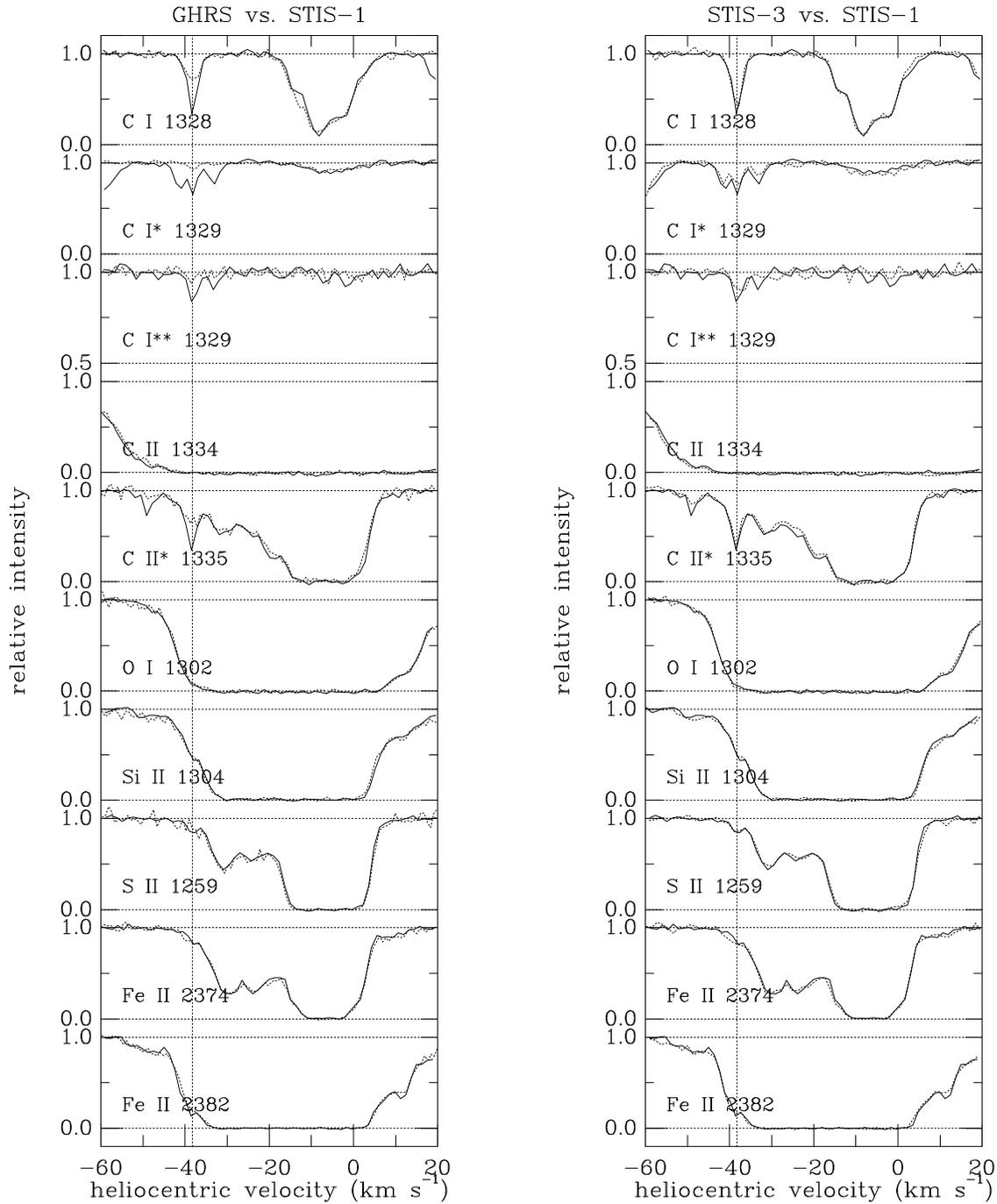}{f2b.eps}
\caption{Profiles of selected UV lines toward HD 219188, observed with the {\it
HST}/GHRS at resolutions of about 3.5 km s$^{-1}$ and with the {\it HST}/STIS at resolutions of about 2.3 km~s$^{-1}$.
The vertical dotted line indicates the IV component at $-$38 km s$^{-1}$.
The left panel compares the GHRS profiles (1994.43/1995.37; {\it dotted lines}) with the STIS-1 profiles (2001.74; {\it solid lines}).
In several cases, different lines observed by GHRS and STIS have been compared by scaling the apparent optical depths of the GHRS profiles by $f\lambda$:  \ion{C}{1} $\lambda$1560 vs. $\lambda$1328, \ion{S}{2} $\lambda$1253 vs. $\lambda$1259, \ion{Fe}{2} $\lambda$2600 vs. $\lambda$2382.
Note the clear increase in the strengths of the trace neutral species C I (both ground and excited states) and of the excited state C II*, while the strengths of the dominant species S II, Si II, and Fe II remained essentially constant.
The right panel compares the STIS-3 profiles (2004.42; {\it dashed lines}) with the STIS-1 profiles (2001.74; {\it solid lines}).
Both C I and C II* appear to have weakened slightly between STIS-1 and STIS-3, but the dominant species again appear to have remained essentially constant.}
\label{fig:h219uv}
\end{figure}

Some of the UV line profiles are shown in Figure~\ref{fig:h219uv}, where the left panel compares GHRS spectra (1994.43, 1995.37) with the first epoch of STIS spectra (2001.74) and the right panel compares the corresponding STIS spectra from the first and third epochs (2001.74, 2004.42).
In both panels, the first STIS spectra (2001.74) are shown by the solid lines.
[For several of the comparisons in the left panel, different lines observed by GHRS and STIS have been compared by scaling the apparent optical depths of the GHRS profiles by $f\lambda$:  \ion{C}{1} $\lambda$1560 vs. $\lambda$1328, \ion{S}{2} $\lambda$1253 vs. $\lambda$1259, \ion{Fe}{2} $\lambda$2600 vs. $\lambda$2382.
Note also that only the strongest of the \ion{C}{1} excited state lines are scaled for the GHRS spectra.]
Inspection of the UV line profiles (and the corresponding apparent optical depth profiles) reveals clear temporal variations near $-$38 km~s$^{-1}$ in the absorption from both the ground-state lines of the trace neutral species \ion{C}{1} and \ion{Mg}{1} and the excited fine-structure levels of \ion{C}{1} and \ion{C}{2}.
No obvious variations are seen at other velocities for those species, however --- or at any velocities for species that are dominant ions in predominantly neutral (\ion{H}{1}) gas.

\subsection{Profile Fitting Analysis}
\label{sec-pf}

Detailed multi-component fits were made to the higher-resolution (KPNO and ESO) \ion{Na}{1} and \ion{Ca}{2} line profiles in order to estimate column densities ($N$), line widths ($b$ $\sim$ FWHM/1.665), and heliocentric velocities ($v$) for the discernible components contributing to the observed absorption (e.g., Welty et al. 2003). 
Profile fits are particularly useful in this case, as they yield column densities for individual blended components, they account explicitly for saturation effects, and they enable comparisons between spectra obtained at different resolutions (coud\'{e} feed vs. CES vs. ARCES; GHRS vs. STIS).
Unless noted otherwise, the rest wavelengths and oscillator strengths for both optical and UV transitions were taken from Morton (2003). 
Simultaneous fits to the ensemble of high-resolution \ion{Ca}{2} profiles required at least 16 components between $-$39 and +6 km~s$^{-1}$; independent fits to the individual high-resolution \ion{Na}{1} profiles required at least nine components, with velocities generally very similar to those found for \ion{Ca}{2} (Table~\ref{tab:comp}).
The adopted component velocities are shown by the tick marks in Figure~\ref{fig:naca}.
The strongest absorption is seen for the low-velocity (LV) components between $-$15 and +3 km~s$^{-1}$, with somewhat weaker absorption at intermediate velocities ($v$ $<$ $-$15 km~s$^{-1}$ and $v$ $>$ +3 km~s$^{-1}$).
As discussed below, the LV and IV components are characterized by somewhat different properties.

The detailed component structures for \ion{Na}{1} and \ion{Ca}{2} then were used to fit the corresponding lower-resolution ARCES line profiles --- fixing the $b$-values and relative velocities for all the components and allowing only the overall velocity zero point and the column density of the $-$38 km~s$^{-1}$ component to vary.
The component velocities obtained for the ARCES spectra thus reflect the (somewhat less well determined) overall velocity zero points of those lower resolution spectra --- and so do not necessarily imply variations in the velocity of the $-$38 km~s$^{-1}$ component.
Moreover, while the $-$38 km~s$^{-1}$ component is reasonably distinct in \ion{Na}{1} in the ARCES spectra, the corresponding \ion{Ca}{2} component lies on the shoulder of the stronger \ion{Ca}{2} absorption at slightly lower (less negative) velocities --- so that the \ion{Ca}{2} column density at $-$38 km~s$^{-1}$ is also less well determined.

The parameters obtained for the IV component near $-$38 km~s$^{-1}$ from all the available \ion{Na}{1} and \ion{Ca}{2} spectra are listed in Table~\ref{tab:naca}; note that some of the values differ slightly from those given by Welty \& Fitzpatrick (2001).
While dramatic variations in column density are seen for the $-$38 km~s$^{-1}$ component, no significant changes in column density were found for any of the other components between $-$32 and +6 km~s$^{-1}$. 
The fits to the lower resolution \ion{Na}{1} and \ion{Ca}{2} profiles thus were obtained under the assumption that only the $-$38 km~s$^{-1}$ component exhibits epoch-to-epoch variations in column density. 
As noted by Welty \& Fitzpatrick (2001), both the velocity ($v$ = $-$38.3$\pm$0.2 km~s$^{-1}$) and the width ($b$ = 0.58$\pm$0.02 km~s$^{-1}$ for \ion{Na}{1}) of the variable IV component appear to be constant, within the uncertainties.
While the $b$-value for \ion{Ca}{2} appears to be slightly larger than that for \ion{Na}{1}, such a difference is not uncommon, and may reflect differences in the spatial distribution of the two species (Welty et al. 1996).

\begin{figure}
\epsscale{1.0}
\plotone{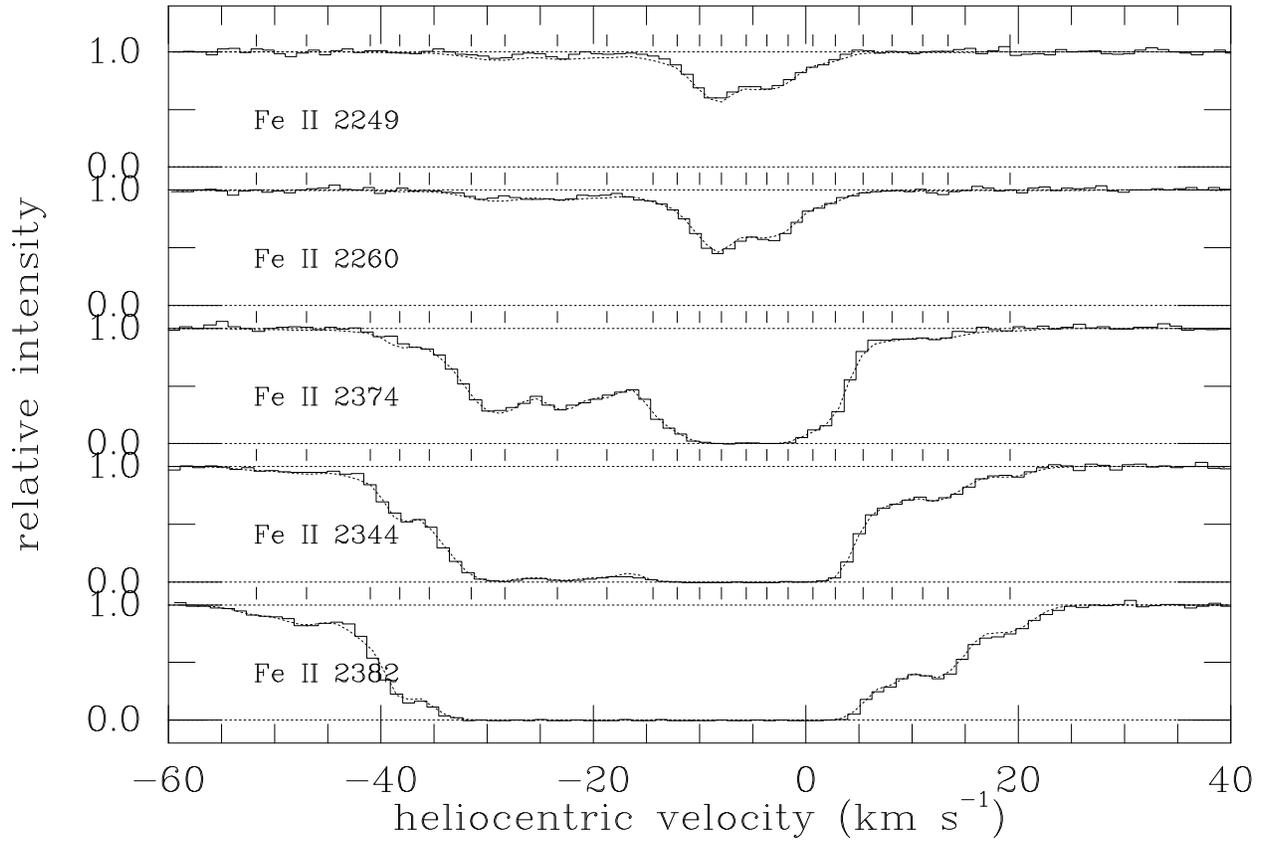}
\caption{Fits to profiles of weak and strong Fe~II lines.
The averaged STIS spectra are given by the solid histograms; simultaneous fits to the profiles are given by the dotted lines (with components noted by tick marks above the spectra).}
\label{fig:h219fit}
\end{figure}

Individual component column densities for the various species detected in the UV spectra were obtained by performing detailed fits to the line profiles, using component structures ($v$ and $b$ for all the discernible components) determined from the highest resolution optical and UV spectra (e.g., Welty et al. 1999b).
As comparisons among the individual STIS spectra of specific transitions revealed significant temporal changes only in the IV component near $-$38 km~s$^{-1}$ (in some species), the STIS spectra from all three epochs were combined to allow more accurate characterization of the other (non-variable) components.
Because the variable IV component near $-$38 km~s$^{-1}$ can be blended with adjacent components within 3--4 km~s$^{-1}$, accurate assessment of the variations requires reliable characterization both of those other IV components and of the overall velocity zero point for each profile.
In the fits to the UV lines, the relative velocities of the LV components were adopted from the fits to the \ion{Na}{1} and \ion{Ca}{2} profiles.
The relative LV column densities for dominant species exhibiting different degrees of depletion were estimated via fits to the relatively weak lines of \ion{O}{1} (1355 \AA), \ion{P}{2} (1302 \AA), \ion{Ge}{2} (1237 \AA), \ion{Mg}{2} (1239, 1240 \AA), \ion{Mn}{2} (1197 \AA), \ion{Fe}{2} (2249, 2260 \AA), and \ion{Ni}{2} (1317 \AA).
Those relative LV column densities then were fixed (i.e., a common scaling factor was applied) in fits to lines showing strong, saturated LV absorption (e.g., the lines of \ion{Si}{2}, \ion{S}{2}, and \ion{N}{1}).
The column densities and relative velocities of the IV components generally were determined via simultaneous fits to several lines covering a range in strength (e.g., the three \ion{Fe}{2} lines at 2344, 2374, and 2382 \AA\ or the three \ion{S}{2} lines at 1250, 1253, and 1259 \AA).
While slight differences in several of the IV component velocities were found in fitting the lines from \ion{Fe}{2}, \ion{S}{2}, and \ion{Mg}{1}, the relative velocities of the two components adjacent to the one at $-$38 km~s$^{-1}$ were held fixed, in order to ``standardize'' the effects of blends with those adjacent components.
As an example, Figure~\ref{fig:h219fit} shows the results of a simultaneous fit (smooth dotted curves) to the average profiles of the five \ion{Fe}{2} lines observed with STIS (solid histograms).
The column densities of the LV components are constrained primarily by the profiles of the weaker lines at 2249 and 2260 \AA\, while the column densities and velocities of the IV components are constrained by the progressively stronger lines at 2374, 2344, and 2382 \AA.
The profiles of lines from the trace species \ion{C}{1}, \ion{Mg}{1} (LV components), and \ion{Cl}{1} were fitted using the structure determined from \ion{Na}{1}; profiles of lines from various dominant species were fitted using the structure determined from \ion{Ca}{2}, \ion{S}{2}, and/or \ion{Fe}{2} (Table~\ref{tab:comp}).
The column densities determined from the average STIS spectra for the $-$38 km~s$^{-1}$ component, the rest of the IV components (at negative velocities), and all the LV components are listed in Table~\ref{tab:ivlv}, together with solar reference abundances from Lodders (2003) and representative depletions for Galactic ``halo'', ``warm, diffuse'' and ``cold, dense'' clouds from Savage \& Sembach (1996), Welty et al. (1999b), Jenkins (2004b), and Cartledge et al. (2006).

Column densities for \ion{C}{1}, \ion{C}{1}*, and \ion{C}{1}** in the $-$38 km~s$^{-1}$ component, obtained from the various \ion{C}{1} multiplets observed with GHRS and STIS (the latter averaged over the three epochs), are given in Table~\ref{tab:c1fs}.
Differences in the column densities derived from the different multiplets may reflect uncertainties in the $f$-values (Jenkins \& Tripp 2001; Morton 2003).
While use of the $f$-values inferred by Jenkins \& Tripp reduces the discrepancies somewhat, they did not analyze all the multiplets observed in this study. 
The adopted \ion{C}{1} column densities --- based on the values obtained from the multiplets at 1560 \AA\ (GHRS) and 1328 \AA\ (STIS), which have similar $f\lambda$ --- are thus somewhat uncertain.
Because the lines from the \ion{C}{1} ground state are typically much stronger than those from the two excited fine-structure levels, the relative column densities in the three levels (used to estimate thermal pressures; see below) are sensitive to the $b$-value used for the $-$38 km~s$^{-1}$ component.
The adopted $b$ = 0.65 km~s$^{-1}$ (which assumes that thermal broadening at $T$ $\sim$ 100 K and identical ``turbulent'' velocities account for the line widths of both \ion{C}{1} and \ion{Na}{1}) yields consistent relative level populations for the multiplets at 1194, 1277, 1280, and 1328 \AA\, which cover a range in $f\lambda$ of nearly an order of magnitude (Table~\ref{tab:c1fs}).

The component structures determined for each species from the average STIS spectra then were used to fit the individual spectra from each epoch, allowing only the column density of the $-$38 km~s$^{-1}$ component and the overall velocity zero point to vary.
The resulting column densities for the $-$38 km~s$^{-1}$ component --- for each epoch with UV spectra --- are listed in Table~\ref{tab:cduv}.
[Note that some of the values derived from the GHRS spectra differ from those given by Welty \& Fitzpatrick (2001), because of slight differences in the component structure adopted in this paper for the IV gas and/or in the $f$-values used for some of the transitions.]
In the table, the values for \ion{Na}{1} and \ion{Ca}{2} (in square brackets) are for the epochs closest to those of the UV data (but within six months).
Because the \ion{C}{2} $\lambda$1334 line is saturated, the values for $N$(\ion{C}{2}) (both ground state and total; in parentheses) are estimated from the observed $N$(\ion{S}{2}) and $N$(\ion{C}{1}), under the assumption of a typical gas-phase interstellar carbon-to-sulfur ratio of 10.4 [which is based on abundances of 161 ppm for carbon (Sofia et al. 2004) and 15.5 ppm for undepleted sulfur (Lodders 2003)].\footnotemark\
\footnotetext{While there may be some S$^{+2}$ associated with the $-$38 km~s$^{-1}$ component, the similar ionization potentials of C$^+$ and S$^+$ suggest that $N$(C~I) + $N$(C~II) may be adequately estimated from $N$(S~I) + $N$(S~II).}

In general, the uncertainties in the column densities include contributions from photon noise, continuum fitting uncertainties, and uncertainties in the adopted $b$-values (which often were held fixed in fitting the individual spectra).
The latter two contributions were estimated via separate fits in which the continua were raised/lowered by a fraction of the observed rms fluctuations and in which the $b$-values were adjusted slightly.
For the $-$38 km~s$^{-1}$ component, the uncertainty in $b$ is most significant for the stronger lines of trace neutral species and \ion{C}{2}* (for which $b$ $\sim$ 0.6--0.7 was required) and for some very strong lines of dominant species (e.g., \ion{O}{1} $\lambda$1302). 
Continuum fitting uncertainties are most significant for column densities derived from weak components in low-to-moderate S/N spectra (e.g., for \ion{S}{2}), but the use of simultaneous fits to multiple lines should reduce those contributions somewhat.
It is harder to quantify the uncertainties in $N$($-$38 km~s$^{-1}$) due to uncertainties in the component structure (e.g., the degree of blending with adjacent components), but the effects should be similar (or at least in the same sense) for all similarly distributed species.
While the column densities of individual IV and LV components can be rather uncertain (due, for example, to uncertainties in component structure and/or in scaling the various components to fit broad, saturated absorption features), most of the total IV and LV column densities appear to be reasonably well determined.

\section{Results}
\label{sec-res}

\subsection{General Properties of Gas toward HD 219188}
\label{sec-gen}

The profiles of the lines of dominant, little-depleted species (e.g., \ion{O}{1} $\lambda$1355, \ion{Mg}{2} $\lambda$1239, \ion{P}{2} $\lambda$1301, \ion{Zn}{2} $\lambda$2062) and of \ion{Cl}{1} indicate that the LV components between $-$15 and +3 km~s$^{-1}$ contain most of the total column densities of \ion{H}{1} and H$_2$ toward HD 219188 (7.0 $\times$ 10$^{20}$ cm$^{-2}$ and 2.2 $\times$ 10$^{19}$ cm$^{-2}$, respectively; Bohlin et al. 1978; Savage et al. 1977).
Average LV column density ratios $N$(\ion{S}{2})/$N$(\ion{S}{3}) $\sim$ 200, $N$(\ion{N}{1})/$N$(\ion{S}{2}) $\sim$ 4, and $N$(\ion{S}{2})/$N$(\ion{O}{1}) $\sim$ 0.06 --- the latter two comparable to values typically found in \ion{H}{1} regions --- suggest that the LV gas is predominantly neutral (Table~\ref{tab:ivlv}).
Average LV values for the ratios $N$(\ion{Fe}{2})/$N$(\ion{S}{2}) $\sim$ 0.05 and $N$(\ion{Ni}{2})/$N$(\ion{S}{2}) $\sim$ 0.003 --- and for the depletions of various elements listed in the next to last column of Table~\ref{tab:ivlv} --- suggest depletions that are generally slightly more severe than those found for ``warm, diffuse'' clouds --- though individual LV components can have depletions that are somewhat more or somewhat less severe.
The most severe depletions, for example, are found for the LV component at $-$10 km~s$^{-1}$, which has the highest column density of \ion{Cl}{1} (and, presumably, of H$_2$).

The IV gas, with components between $-$52 and $-$18 km~s$^{-1}$, has a significantly smaller total $N$(H) = $N$(\ion{H}{1}) + $N$(\ion{H}{2}) $\sim$ 1.7 $\times$ 10$^{19}$ cm$^{-2}$ (estimated from the column densities of \ion{S}{2} and \ion{S}{3} with the assumption that sulfur is undepleted).
Lower values for the ratios $N$(\ion{S}{2})/$N$(\ion{S}{3}) $\sim$ 3 and $N$(\ion{N}{1})/$N$(\ion{S}{2}) $\sim$ 0.5 suggest that the IV gas (on average) is partially ionized.
The $N$(\ion{Mg}{1})/$N$(\ion{Na}{1}) ratios, higher by more than an order of magnitude than in the LV components (except in the $-$38 km~s$^{-1}$ component), suggest that most of the IV gas is relatively warm ($T$ $\ga$ 5000 K), with the \ion{Mg}{1} enhanced primarily via dielectronic recombination (e.g., Welty et al. 1999a).
Higher values of $N$(\ion{Fe}{2})/$N$(\ion{S}{2}) $\sim$ 0.5 and $N$(\ion{Ni}{2})/$N$(\ion{S}{2}) $\sim$ 0.03 indicate that the IV depletions generally are milder than those in the LV components. 
The variable IV component at $-$38 km~s$^{-1}$ --- which exhibits relatively strong absorption from several trace neutral species, \ion{Ca}{2}, and \ion{C}{2}* in cooler gas ($T$ $\la$ 500 K) --- thus differs somewhat from the other IV components.

\subsection{Temporal Variations in $-$38 km~s$^{-1}$ Component}
\label{sec-tvar}

\begin{figure}
\epsscale{1.0}
\plotone{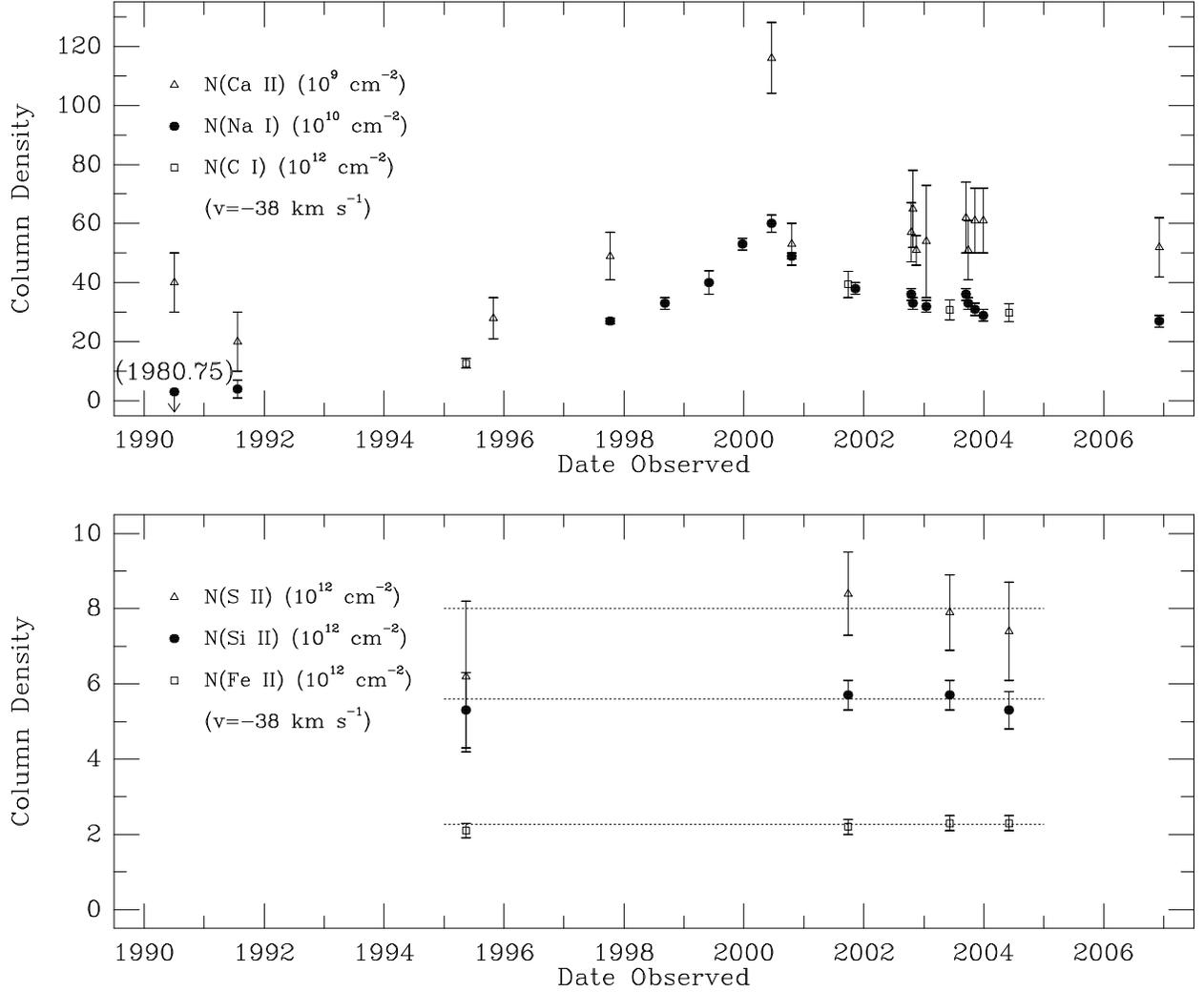}
\caption{Variations in $N$(Ca II), $N$(Na I), and $N$(C I) ({\it top}) and in $N$(S II), $N$(Si II), and $N$(Fe II) ({\it bottom}), at $v$ $\sim$ $-$38 km s$^{-1}$ toward HD~219188.
Na I was not detected, and Ca II only weakly, in 1980 (Albert 1983); a weak Na I feature may have been detected in 1991 (Sembach et al. 1993).
$N$(Na I) ({\it filled circles}) increased steadily between 1997.77 and 2000.46, but then decreased between 2000.46 and 2006.93; the changes in $N$(C I) ({\it open squares}) and $N$(Ca II) ({\it open triangles}) generally appear to parallel those in $N$(Na I).
$N$(S II), $N$(Si II), and $N$(Fe II) --- and thus $N$(H) and the (mild) depletions --- appear to have remained essentially constant between 1995.37 and 2004.42; the dotted lines in the bottom panel show the mean values for the three epochs with STIS data.}
\label{fig:h219var}
\end{figure}

Between 1997.77 and 2000.46, the column density of \ion{Na}{1} in the IV component at $-$38 km~s$^{-1}$ increased from $\sim$ 3 $\times$ 10$^{11}$ cm$^{-2}$ (an order of magnitude higher than the limit found in 1980) to $\sim$ 6 $\times$ 10$^{11}$ cm$^{-2}$ (Table~\ref{tab:naca} and Fig.~\ref{fig:h219var}).  
By the end of 2003, however, $N$(\ion{Na}{1}) decreased again to $\sim$ 3 $\times$ 10$^{11}$ cm$^{-2}$, then continued to decrease (slowly) through the end of 2006.
Similar variations are seen both for $N$(\ion{C}{1}) and $N$(\ion{C}{2}*) (which increased by factors of 3--4 between 1995.37 and 2001.74, then decreased by 20--40\% by 2004.42) and for $N$(\ion{Ca}{2}), except for a brief period in late 2000 in which $N$(\ion{Ca}{2}) appears to have declined more rapidly than $N$(\ion{Na}{1}).
The column density of \ion{Mg}{1}, detected at $-$38 km~s$^{-1}$ only in the STIS spectra, shows a similar weak decrease between 2001.74 and 2004.42.
The $N$(\ion{Na}{1})/$N$(\ion{Ca}{2}) ratio thus remained roughly constant, at a value 6--8 times higher than the limit found in 1980.
Variations are also seen in the relative fine-structure populations of \ion{C}{1} (see below).
If the variations in the trace neutral species, \ion{Ca}{2}, and \ion{C}{2}* are due solely to the proper motion of HD~219188, the roughly 5-year ``FWHM'' corresponds to a transverse linear scale of 10--200 AU (depending on the distance to the cloud).

The column densities of the various dominant species (e.g., \ion{S}{2}, \ion{Si}{2}, \ion{Fe}{2}), however, exhibited no significant variations between 1994--1995 and 2004 (Table~\ref{tab:cduv} and Fig.~\ref{fig:h219var}).
While the nominal values for $N$(\ion{S}{2}) would be consistent with a slight ($\sim$30\%) increase between 1995.37 and 2001.74, then smaller decreases for 2003.43 and 2004.42, the $-$38 km~s$^{-1}$ component is relatively weak and those ``variations'' are within the 1 $\sigma$ uncertainties.
The column densities of \ion{Si}{2} and \ion{Fe}{2} --- which are derived from somewhat stronger lines --- have smaller relative uncertainties and show no apparent variations.

\subsection{Physical Properties of $-$38 km~s$^{-1}$ Component}
\label{sec-phys}

\subsubsection{$N$(H) and Depletions}
\label{sec-nhdep}

If sulfur is undepleted in the $-$38 km~s$^{-1}$ component, then the sum of the column densities measured from the averaged STIS spectra for \ion{S}{2} and \ion{S}{3}, 9.3 $\times$ 10$^{12}$ cm$^{-2}$, would imply a total hydrogen column density $N$(\ion{H}{1}) + $N$(\ion{H}{2}) of 6.0 $\times$ 10$^{17}$ cm$^{-2}$, or about 3\% of the total hydrogen in all the IV components (Table~\ref{tab:ivlv}).
A similar estimate for $N$(\ion{H}{1}) based on the observed $N$(\ion{O}{1}) (assumed to be depleted by 0.1 dex) is about 2.2 $\times$ 10$^{17}$ cm$^{-2}$ --- which would suggest that roughly 65\% of the hydrogen in the $-$38 km~s$^{-1}$ component is ionized (but with large uncertainty).
Estimates for the electron and hydrogen densities in the $-$38 km$^{-1}$ component (see next sections), however, imply that the hydrogen is ``only'' about 10--20\% ionized.
As \ion{Cl}{1} is enhanced when H$_2$ is abundant (e.g., Jura \& York 1978), the upper limit on the $N$(\ion{Cl}{1})/$N$(\ion{Na}{1}) ratio in the $-$38 km~s$^{-1}$ component --- much lower than the value found for the LV components --- indicates that there is very little H$_2$ in the $-$38 km~s$^{-1}$ component.
The column density ratios $N$(\ion{Si}{2})/$N$(\ion{S}{2}) $\sim$ 0.7 and $N$(\ion{Fe}{2})/$N$(\ion{S}{2}) $\sim$ 0.3 --- similar to those found for the other IV components and for clouds in the Galactic halo (Tables~\ref{tab:ivlv} and \ref{tab:cduv}) --- suggest depletions that are less severe than those found for the LV components.\footnotemark\
The lack of significant variation in the column densities of the various dominant species implies that the total $N$(H) and the relatively mild depletions remained essentially unchanged between 1995 and 2004 [although the brief rise in $N$(\ion{Na}{1})/$N$(\ion{Ca}{2}) in late 2000 may be indicative of a small region with slightly more severe depletions].
\footnotetext{Depletions are estimated relative to \ion{S}{2}, since the $-$38 km~s$^{-1}$ gas is partially ionized but column densities are generally not available for all potentially significant ionization stages (e.g., for \ion{Fe}{3}); see discussions in Welty et al. 1997).}

\subsubsection{Thermal Pressure and Local Hydrogen Density}
\label{sec-nh}

Estimates of the local thermal pressure $n_{\rm H}T$ in relatively cool, predominantly neutral interstellar clouds may be obtained from the relative populations in the excited fine-structure levels of \ion{C}{1} (Jenkins \& Shaya 1979; Jenkins \& Tripp 2001).
If an estimate for the temperature is available, then the local hydrogen density $n_{\rm H}$ (including both \ion{H}{1} and H$_2$) may also be determined.
Because of uncertainties in the \ion{C}{1} oscillator strengths (see above), the relative populations $f_1$ = $N$(\ion{C}{1}*)/$N$(\ion{C}{1}$_{\rm tot}$) and $f_2$ = $N$(\ion{C}{1}**)/$N$(\ion{C}{1}$_{\rm tot}$) were determined from individual \ion{C}{1} multiplets instead of from a global fit to all the multiplets.
For the three STIS epochs, consistent values for the relative populations were obtained from the four \ion{C}{1} multiplets at 1194, 1277, 1280, and 1328 \AA --- which cover a range in $f\lambda$ of nearly a factor of 10 and whose lines are not blended with lines from other species (Table~\ref{tab:c1fs}).
Only the multiplets at 1193, 1194, and 1560 \AA\ were observed with GHRS.
The \ion{C}{1}, \ion{C}{1}*, and \ion{C}{1}** column densities adopted for the $-$38 km~s$^{-1}$ component for each of the four epochs are listed in Table~\ref{tab:cduv}; the corresponding relative fine-structure populations $f_1$ and $f_2$ are given in Table~\ref{tab:rat}. 
The higher values for $f_1$ derived from the STIS spectra are indicative of higher thermal pressures and densities in the $-$38 km~s$^{-1}$ component during 2001--2004 (compared to 1995); $f_1$ (and thus $n_{\rm H}T$) may have declined slightly between 2001.74 and 2004.42.

The derived relative \ion{C}{1} fine-structure populations are plotted ($f_2$ vs. $f_1$) in Figure~\ref{fig:h219c1fs} together with theoretical curves for gas (at $T$ = 50, 100, and 200 K) exposed to the WJ1 interstellar radiation field (de Boer et al. 1973) and characterized by 10\% ionization of hydrogen (see next section).
The predicted fine-structure populations were calculated using collisional rates from Launay \& Roueff (1977) for H atoms, from Roueff \& Le Bourlot (1990) for protons, from Staemmler \& Flower (1991) for He$^0$, and from Johnson et al. (1987) for electrons (see, e.g., Jenkins \& Tripp 2001).\footnotemark\
The open squares along each of the curves mark the positions corresponding to log[$n_{\rm H}$ (cm$^{-3}$)] = 0, 1, 2, 3, and 4 (increasing away from the origin).
If we adopt $T$ = 100 K, consistent with the limits imposed by \ion{C}{2} fine-structure excitation ($T$ $>$ 50 K; see next section) and $b$(\ion{Na}{1}) ($T$ $<$ 490 K), then the GHRS data imply $n_{\rm H}T$ $\sim$ 2000 cm$^{-3}$K and $n_{\rm H}$ $\sim$ 20 cm$^{-3}$ in 1995, while the STIS data imply $n_{\rm H}T$ $\sim$ 4500--3400 cm$^{-3}$K and $n_{\rm H}$ $\sim$ 45--34 cm$^{-3}$ in 2001--2004 (Table~\ref{tab:phys}).
(The thermal pressures are not very sensitive to the exact value of $T$, however, and are within the listed uncertainties for both 50 K and 200 K.)
Together with the $N$(H) estimated above, these local hydrogen densities imply thicknesses of order 1000--1600 AU for the $-$38 km~s$^{-1}$ component.
The thermal pressure in the $-$38 km~s$^{-1}$ component is somewhat higher than those in the predominantly neutral LV components, which (on average) have $n_{\rm H}T$ $\sim$ 600 cm$^{-3}$K.
The relative populations determined from the STIS data for the $-$38 km~s$^{-1}$ component fall very close to the theoretical curves --- suggesting that that component could represent gas at a single, uniform pressure.
\footnotetext{The program employed for those calculations was developed (over many years) by P. Frisch, D. Welty, and J. Lauroesch.}

Because the \ion{C}{1} excited fine-structure levels may also be populated via photon pumping, the relative populations may also be used to derive constraints on the strength of the ambient radiation field (e.g., Jenkins \& Shaya 1979; Lauroesch et al. 1998).
For the $-$38 km~s$^{-1}$ component toward HD~219188, the relative populations derived from the STIS data ($f_1$ = 0.26--0.30) indicate that the radiation field cannot be stronger than about 40 times the WJ1 field --- which suggests that the gas must be at least several pc away from the star.

\subsubsection{Electron Density and Fractional Ionization}
\label{sec-ne}

The relative prominence of the $-$38 km~s$^{-1}$ component in lines from the trace neutral species \ion{C}{1}, \ion{Na}{1}, and \ion{Mg}{1} and from \ion{C}{2}* is suggestive of a relatively high electron density in that component.
Estimates for $n_e$ may be obtained both from trace/dominant ratios (e.g., \ion{C}{1}/\ion{C}{2}) under the (questionable) assumption of photoionization equilibrium and from analysis of the \ion{C}{2} fine-structure excitation.
If we again adopt $T$ = 100 K and the WJ1 interstellar radiation field (de Boer et al. 1973), then the ratio of photoionization and recombination rates $\Gamma$/($\alpha~n_e$) for carbon is about 24/$n_e$ (P\'{e}quignot \& Aldrovandi 1986).
In photoionization equilibrium, the electron density, $n_e$ = 24~$N$(\ion{C}{1})/$N$(\ion{C}{2}), would then have been about 6 cm$^{-3}$ in 1995 (GHRS A), about 20 cm$^{-3}$ in 2001 (STIS 1), and about 15 cm$^{-3}$ in 2003 and 2004 (STIS 2, STIS 3) (Table~\ref{tab:phys}). 
Together with the local hydrogen densities derived from \ion{C}{1} fine-structure excitation, these electron densities would imply fractional ionizations $n_e$/$n_{\rm H}$ of about 30\% in 1995 and about 40--45\% in 2001--2004 --- in rough (and perhaps fortuitous) agreement with the fractional ionization of hydrogen estimated from the column densities of \ion{O}{1}, \ion{S}{2}, and \ion{S}{3} (\S\ref{sec-nhdep}).
These estimates for $n_e$ and $n_e$/$n_{\rm H}$ are much larger than the values typically found for diffuse neutral gas in sight lines with much higher $N$(H), where $n_e$ $\sim$ 0.05--0.20 cm$^{-3}$ and $n_e$/$n_{\rm H}$ is usually less than 1\% (e.g., Welty et al. 2003).
As the recombination coefficient $\alpha$ is proportional to $T^{-0.62}$, the estimates for $n_e$ and $n_e$/$n_{\rm H}$ would be about 35\% lower for $T$ = 50 K and about 55\% higher for $T$ = 200 K.
Both $n_e$ and $n_e$/$n_{\rm H}$ would be proportionally lower if the radiation field were weaker than the assumed WJ1 field.
While the limits on the ratios $N$(\ion{S}{1})/$N$(\ion{S}{2}) and $N$(\ion{Fe}{1})/$N$(\ion{Fe}{2}) yield corresponding limits on $n_e$ that are significantly smaller than the values estimated from $N$(\ion{C}{1})/$N$(\ion{C}{2}), such differences have been seen in other sight lines (Welty et al. 1999b, 2003).

An alternative (and probably more reliable) estimate for the electron density may be obtained from analysis of the fine-structure excitation of \ion{C}{2} (e.g., Fitzpatrick \& Spitzer 1997; Welty et al. 1999b).
For the temperature and densities characterizing the $-$38 km~s$^{-1}$ component, the population of the excited state \ion{C}{2}* is determined primarily by electron collisional excitation and radiative decay, with a small contribution from collisions with hydrogen atoms.
If we adopt $T$ = 100 K, the $n_{\rm H}$ estimated from \ion{C}{1} fine-structure excitation (above), and the collisional and radiative rates used by Welty et al. (1999b), then the column density ratios $N$(\ion{C}{2}*)/$N$(\ion{C}{2}) listed in Table~\ref{tab:rat} yield electron densities of about 0.7 cm$^{-3}$ in 1995 (GHRS A), about 8 cm$^{-3}$ in 2001 (STIS 1), and about 3--5 cm$^{-3}$ in 2003--2004 (STIS 2, STIS 3) --- factors of 2.5--8.5 lower than the values inferred from photoionization equilibrium of carbon.
The corresponding fractional ionizations are about 0.035 for 1995 and about 0.07--0.17 for 2001--2004.
The estimates for $n_e$ and $n_e$/$n_{\rm H}$ would be slightly lower for $T$ = 200 K; the $N$(\ion{C}{2}*)/$N$(\ion{C}{2}) ratios obtained from the STIS data are inconsistent with temperatures below about 50 K, however.

It is not uncommon for estimates of $n_e$ obtained from trace/dominant ratios under the assumption of photoionization equilibrium to be somewhat higher than those from analysis of \ion{C}{2} fine-structure excitation (e.g., Fitzpatrick \& Spitzer 1997; Welty et al. 1999b).
One possible explanation for such differences is that the abundances of the trace neutral species are enhanced by charge exchange between the dominant first ions and neutral or negatively charged large molecules/small grains (Lepp et al. 1988; Welty \& Hobbs 2001; Weingartner \& Draine 2001; Liszt 2003; Welty et al. 2003).
Because $N$(\ion{C}{1})/$N$(\ion{C}{2}) is large and $n_{\rm H}$ is relatively small, however, such effects should not be not significant in this case (Weingartner \& Draine 2001; Welty et al. 2003).
As noted above, the differences in inferred $n_e$ could also be reduced if the $-$38 km~s$^{-1}$ component is subject to a radiation field that is weaker than the assumed WJ1 field.
Even the lower fractional ionizations determined from \ion{C}{2} fine-structure excitation, however, are greater than the value $\sim$ 0.00016 that would correspond to the electrons being due primarily to ionization of carbon (as is usually assumed) --- implying that hydrogen must be partially ionized.
While the uncertainties in the $n_e$ estimated in both ways are dominated by the uncertainties in $N$(\ion{C}{2}) [which is estimated from the measured $N$(\ion{S}{2}) and $N$(\ion{C}{1}$_{\rm tot}$) and an assumed gas-phase ratio of sulfur to (total) carbon], the ratio of the two $n_e$ estimates is relatively insensitive to $N$(\ion{C}{2}).
The increased column densities of both the trace neutral species and \ion{C}{2}* determined from the STIS spectra [together with the roughly constant $N$(\ion{X}{2})] thus indicate that both $n_e$ and $n_e$/$n_{\rm H}$ increased by factors of a few between 1995 and 2001--2004.

\section{Discussion}
\label{sec-disc}

\subsection{Are Temporal Variations Associated with High Pressures?}
\label{sec-highp}

In most cases where small-scale spatial or temporal variations have been observed in interstellar \ion{Na}{1}, there is little direct information on the physical conditions in the absorbing gas (e.g., from UV spectra), and the presence of at least some relatively dense gas ($n_{\rm H}$ $\ga$ 10$^3$ cm$^{-3}$) has commonly been inferred (often on the basis of an assumed \ion{Na}{1}/H ratio).
In the few sight lines with UV data, however, analysis of the \ion{C}{1} fine-structure excitation suggests that the thermal pressures and densities in the bulk of the gas exhibiting the variations are actually relatively low.
Toward the binary system $\mu^1$, $\mu^2$ Cru, the trace neutral species \ion{C}{1}, \ion{Na}{1}, and \ion{Mg}{1} are enhanced in a narrow component at $-$8.6 km~s$^{-1}$ toward $\mu^1$ (Lauroesch et al. 1998).
The dominant ions \ion{Cr}{2} and \ion{Zn}{2} show no significant differences at that velocity, however, and the non-detection of the excited state lines of \ion{C}{1} indicates a fairly low density. 
Toward HD 32040, Lauroesch et al. (2000) observed temporal variations in \ion{Na}{1} in a component at 9 km~s$^{-1}$, but the upper limits on corresponding absorption from \ion{C}{1}* imply that $n_{\rm H}$ $<$ 26 cm$^{-3}$ (for $T$ = 100 K).
Comparisons between the absorption from the trace neutral species and from \ion{Zn}{2} suggest that hydrogen is slightly ($\sim$ 1\%) ionized in that component.
Toward $\rho$ Leo, Lauroesch \& Meyer (2003) found variations in the moderately strong, narrow, isolated component at 18 km~s$^{-1}$, but the non-detection of \ion{C}{1}* at that velocity again yields a fairly low limit on the density ($n_{\rm H}$ $<$ 20 cm$^{-3}$).
That 18 km~s$^{-1}$ component appears to be characterized by ``warm cloud'' depletions and some slight ionization of hydrogen.
The densities found here for the $-$38 km~s$^{-1}$ component toward HD~219188, $n_{\rm H}$ $\sim$ 20--45 cm$^{-3}$, are slightly higher than the limits found for the variable components toward HD~32040 and $\rho$ Leo, but they are nonetheless much lower than the values estimated by assuming a ``typical'' \ion{Na}{1}/H ratio (see below).

\begin{figure}
\epsscale{0.7}
\plotone{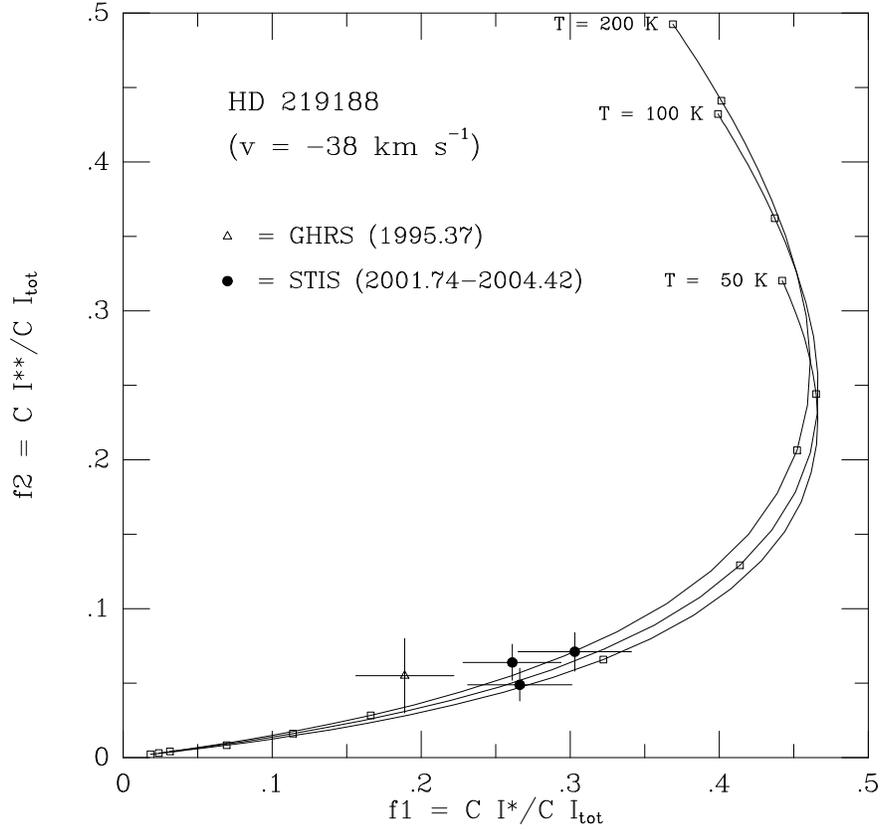}
\caption{C I fine-structure excitation in the $-$38 km~s$^{-1}$ component toward HD~219188.
The GHRS value (1995.37) is given by the open triangle; the three STIS values (2001.74--2004.42) are given by the filled circles.
The solid curves show the predicted ratios for $T$ = 50, 100, and 200 K (Jenkins \& Shaya 1979; Jenkins \& Tripp 2001), assuming the WJ1 interstellar radiation field and 10\% ionization of hydrogen; the open squares along each curve denote log($n_{\rm H}$) = 0, 1, 2, 3, 4.
The three STIS values are in good agreement with the predicted curves --- suggesting that the IV gas may be characterized by uniform pressure (i.e., not a mixture of low- and high-pressure gas).}
\label{fig:h219c1fs}
\end{figure}

While the analyses of \ion{C}{1} fine-structure excitation in those few sight lines indicate that most of the gas in the variable components is at relatively low pressures and densities, the relative fine-structure populations derived from the GHRS spectra for the $-$38 km~s$^{-1}$ component toward  HD~219188 (Welty \& Fitzpatrick 2001) lie slightly above the theoretical curves in Figure~\ref{fig:h219c1fs} --- consistent with the presence of a small amount of much denser gas (Jenkins \& Tripp 2001; Crawford 2003). 
Lauroesch \& Meyer (2003), noting that the trace neutral species preferentially sample denser gas, proposed that the observed variable components might be due to small regions of dense, high-pressure gas --- similar to those inferred for many sight lines in the wider \ion{C}{1} survey of Jenkins \& Tripp (2001).
Such small, high-density regions might result from interstellar turbulence (e.g., Jenkins 2004a).
Lauroesch \& Meyer then suggested that one possible consequence of that picture would be that the length scale for the fluctuations of a given trace neutral species should depend on the photoionization rate --- so that \ion{C}{1} (with $\Gamma$ = 20 $\times$ 10$^{-11}$ s$^{-1}$ for the WJ1 field; P\'{e}quignot \& Aldrovandi 1986) would exhibit larger variations than \ion{Na}{1} (with $\Gamma$ = 1.3 $\times$ 10$^{-11}$ s$^{-1}$).
As noted above, however, the more precise \ion{C}{1} populations for the $-$38 km~s$^{-1}$ component determined from the STIS data for HD~219188 fall very close to the theoretical curves --- so that that variable component could be due to gas at a single (relatively low) pressure and density.
The weak \ion{C}{1}** absorption inferred from the GHRS data --- which in any case yields an $f_2$ less than 1.5 $\sigma$ away from the theoretical curves (Fig.~\ref{fig:h219c1fs}) --- may have been slightly overestimated.
Furthermore, the variations observed for \ion{C}{1} during 2001--2004 are very similar to those seen for \ion{Na}{1} (Fig.~\ref{fig:h219var}) --- so that while the \ion{C}{1}/\ion{Na}{1} ratio in the $-$38 km~s$^{-1}$ component is somewhat higher than usual (e.g., Welty \& Hobbs 2001), it also appears to be roughly constant.

\subsection{Relationship between $N$(Na~I) and $N$(H)}
\label{sec-nah}

\begin{figure}
\epsscale{0.7}
\plotone{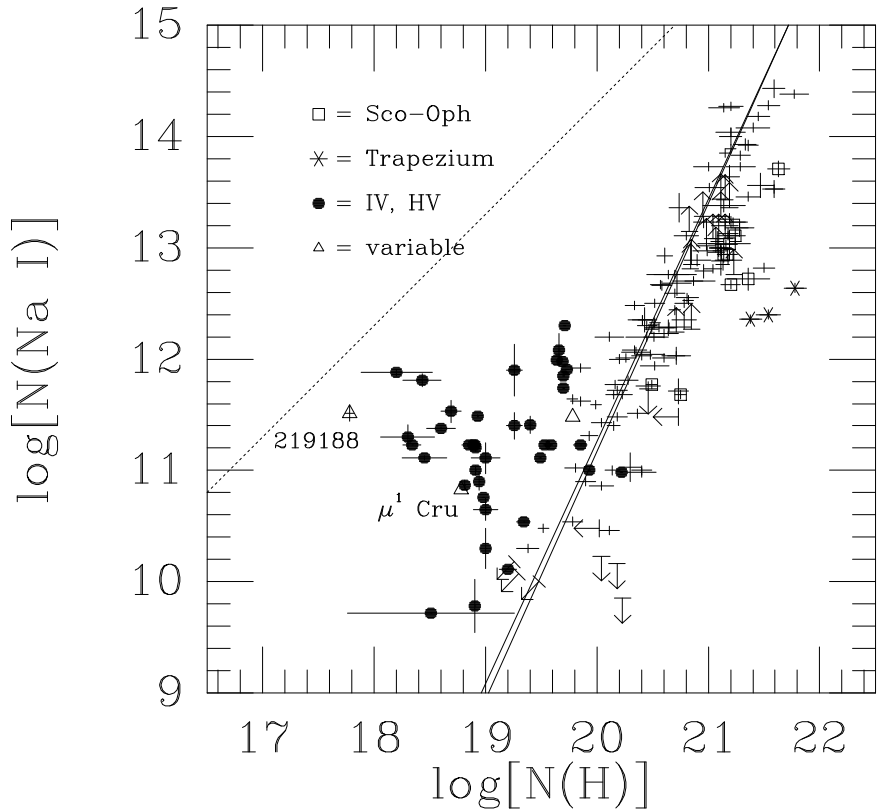}
\caption{$N$(Na I) vs. $N$(H) for sight lines (mostly) in the local Galactic ISM (e.g., Welty \& Hobbs 2001).
The solid lines, with slopes $\sim$ 2.2, are weighted and unweighted least-squares fits to the Galactic data; the dotted line shows the solar abundance of sodium (Lodders 2003).
Na I is somewhat deficient for sight lines in the Sco-Oph and Orion Trapezium regions ({\it open squares} and {\it asterisks}, respectively), but appears to be enhanced for the variable components in two of the three sight lines (HD~219188, $\mu^1$ Cru, $\rho$ Leo; {\it open triangles}) for which $N$(H) can be estimated.
In particular, the IV component toward HD 219188 has a very high $N$(Na I)/$N$(H) ratio --- more similar to the values found for some other IV or high-velocity components ({\it filled circles}; Wakker 2001).
$N$(H II) is not known for those latter components, however.}
\label{fig:h219nah}
\end{figure}

In the local Galactic ISM, there is a fairly well defined, roughly quadratic relationship between the column densities of \ion{Na}{1} and (total) hydrogen for $N$(\ion{Na}{1}) $\ga$ 2--3 $\times$ 10$^{11}$ cm$^{-2}$ and $N$(H) $\ga$ 10$^{20}$ cm$^{-2}$ (with some regional differences), but there is considerable scatter (or perhaps a different relationship?) at lower column densities (Welty et al. 1994; Wakker \& Mathis 2000; Welty \& Hobbs 2001).
Those trends are evident in Figure~\ref{fig:h219nah}, where the crosses show total column densities for a sample of Galactic sight lines (Welty \& Hobbs 2001), the asterisks and open squares show the generally lower $N$(\ion{Na}{1}) seen in a number of sight lines in the Sco-Oph and Orion Trapezium regions, and the filled circles show values for individual intermediate- or high-velocity components from the compilation of Wakker (2001).
The diagonal dotted line shows the solar abundance of sodium (Lodders 2003).
For the Galactic sight lines, $N$(H) is given by the sum of the column densities of \ion{H}{1} and H$_2$ (both derived from UV absorption lines) --- which should be reasonably accurate for values greater than about 10$^{20}$ cm$^{-2}$.
For the intermediate- and high-velocity components [most with $N$(H) $<$ 10$^{20}$ cm$^{-2}$], however, $N$(H) is actually just the $N$(\ion{H}{1}) derived from 21 cm emission, and there may well be some (unknown) amount of \ion{H}{2} present as well --- as appears to be the case for the $-$38 km~s$^{-1}$ component toward HD~219188. 

The assumption that a similar relationship holds for individual components that vary in $N$(\ion{Na}{1}) (either spatially or temporally) may be tested via measurements of UV lines from dominant, little depleted species such as \ion{S}{2} and \ion{Zn}{2}, which may be used to estimate $N$(H).
The resulting points for the $-$8.6 km~s$^{-1}$ component toward $\mu^1$ Cru (Lauroesch et al. 1998), the 18 km~s$^{-1}$ component toward $\rho$ Leo (Lauroesch \& Meyer 2003), and the $-$38 km~s$^{-1}$ component toward HD~219188 (this paper; Table~\ref{tab:phys}) are shown by open triangles in Figure~\ref{fig:h219nah}.
While the point for the 18 km~s$^{-1}$ component toward $\rho$ Leo [near (19.8,11.5)] falls fairly close to the average local Galactic relationship, the variable components toward $\mu^1$ Cru [near (18.8,10.8)] and especially HD~219188 [near (17.8,11.5)] have much lower $N$(H) than would be predicted using that average relationship.
Those latter two points are, however, consistent with those found for the individual intermediate- and high-velocity components.
Estimates of $N$(H) (and corresponding pressures and densities) based on the $N$(\ion{Na}{1}) observed in other sight lines exhibiting small-scale spatial or temporal variability must thus be viewed as highly uncertain --- particularly where $N$(\ion{Na}{1}) is less than 10$^{12}$ cm$^{-2}$.

\subsection{Other Low-$N$(H) Clouds}
\label{sec-lownh}

The $N$(H) inferred for the $-$38 km~s$^{-1}$ component toward HD~219188, about 6 $\times$ 10$^{17}$ cm$^{-2}$, is similar to the values found for some thin, cold Galactic clouds recently found in sensitive 21 cm observations of extragalactic continuum sources (Braun \& Kanekar 2005; Stanimirovi\'{c} \& Heiles 2005; Stanimirovi\'{c} et al. 2007). 
Stanimirovi\'{c} et al. conjecture that these clouds, with $N$(\ion{H}{1}) $\sim$ 10$^{18}$ cm$^{-2}$, may represent the low column density end of a broad distribution of cold, neutral clouds --- at column densities much lower than those expected both from theoretical models of the ISM (e.g., McKee \& Ostriker 1977) and from previous 21 cm emission/absorption observations (e.g., Heiles \& Troland 2003); they may be numerous enough to contain a few percent of the total interstellar \ion{H}{1}.
If these clouds are characterized by typical interstellar thermal pressures, then their densities would lie in the range from about 20--100 cm$^{-3}$ --- very similar to the values found for the variable IV component toward HD~219188.
Stanimirovi\'{c} \& Heiles (2005) briefly discuss several possible scenarios that could account for the existence of such thin clouds, including 
(1) the creation of relatively long-lived small cold clouds in the collisions of turbulent flows (Audit \& Hennebelle 2005; see also Hennebelle \& Audit 2007; Hennebelle et al. 2007);
(2) the generation of transient, relatively low-density clouds (also) as a result of interstellar turbulence (e.g., Vazquez-Semadini et al. 1997), and
(3) the formation of small cloud fragments via shock wave disruption of larger clouds.
In some of those scenarios, the small, cold clouds are surrounded by warmer, more extensive neutral gas, which can help shield the cold gas from ionization (and thus prolong its existence).
Toward HD~219188, the other IV components appear to be warm, partially ionized, and (in most cases) characterized by somewhat higher $N$(H) than the colder variable component at $-$38 km~s$^{-1}$. 
Differences in velocity of at least 3--4 km~s$^{-1}$, however, suggest that there is no clear candidate among those other IV components that could provide such a protective envelope around the colder gas.

\section{Summary}
\label{sec-summ}

Since 1980, the sight line toward the low halo star HD~219188 has exhibited dramatic variations in interstellar \ion{Na}{1} absorption in an intermediate-velocity component at $-$38 km~s$^{-1}$.
If those variations are due solely to the 13 mas yr$^{-1}$ proper motion of HD~219188, they probe spatial scales of 2--38 AU yr$^{-1}$ in the corresponding interstellar cloud at that velocity.
In this paper, we have discussed multi-epoch optical and UV spectra of HD~219188, which were obtained between 1994 and 2006 in order to monitor the variations and characterize the physical conditions in that interstellar cloud.

Optical spectra of interstellar \ion{Na}{1} and/or \ion{Ca}{2}, obtained at resolutions of 1.2--8.0 km~s$^{-1}$ at several different facilities, are now available for 19 epochs between 1980.75 and 2006.93.
The column density of \ion{Na}{1}, which was undetected at $-$38 km~s$^{-1}$ in 1980, increased by a factor of at least 20 by the end of 2000 (to a maximum value $\sim$ 6 $\times$ 10$^{11}$ cm$^{-2}$), then declined by a factor $\ga$ 2 by the end of 2006. 
The roughly 5 year ``FWHM'' of the variations corresponds to a transverse spatial scale of 10--200 AU.
The narrow \ion{Na}{1} line width ($b$ = 0.58 km~s$^{-1}$) implies a temperature less than 490 K for that IV cloud.
Within the uncertainties, both the width and the velocity of the $-$38 km~s$^{-1}$ component (as well as the column densities of the various other components along the sight line) appear to have remained constant over the observed period.
Between 1997 and 2006, the column density of \ion{Ca}{2} exhibited variations similar (in general) to those seen for \ion{Na}{1} --- so that the ratio $N$(\ion{Na}{1})/$N$(\ion{Ca}{2}) also remained roughly constant (although at a value somewhat higher than the limit found in 1980).

UV spectra of HD~219188, covering lines from a number of species, have been obtained with the {\it HST}/GHRS (1994.43, 1995.37; FWHM $\sim$ 3.5 km~s$^{-1}$) and with the {\it HST}/STIS (2001.74, 2003.43, 2004.42; FWHM $\sim$ 2.3 km~s$^{-1}$).
Comparisons between the column densities of \ion{O}{1}, \ion{S}{2}, and \ion{S}{3} suggest that the gas at $-$38 km~s$^{-1}$ is partially ionized.
No significant variations are seen for the column densities of species that are dominant in such gas (e.g., \ion{S}{2}, \ion{Si}{2}, \ion{Fe}{2}) --- suggesting that both the inferred hydrogen column density [$N$(H) $\sim$ 6 $\times$ 10$^{17}$ cm$^{-2}$] and the relatively mild depletions in the $-$38 km~s$^{-1}$ component were essentially unchanged between 1994 and 2004.
The column densities of the trace neutral species \ion{C}{1} (and its excited fine-structure states) and of the excited state \ion{C}{2}*, however, increased by factors of 2--5 between 1995 and 2001, then declined slightly through 2003 and 2004 --- behavior roughly similar to that of $N$(\ion{Na}{1}) over that period.
The hydrogen column densities inferred from the UV data are much lower than the values that would be estimated under the assumption of ``typical'' $N$(\ion{Na}{1})/$N$(H) ratios --- suggesting that \ion{Na}{1}-based estimates of $N$(H) (and corresponding pressures and densities) made for spatially or temporally variable components in other sight lines should be viewed as very uncertain. 

The variations in the \ion{C}{1} fine-structure excitation imply that the thermal pressure $n_{\rm H}T$ in the $-$38 km~s$^{-1}$ component increased by a factor of about 2 between 1995 and 2001, then declined slightly in 2003--2004.
The relative \ion{C}{1} excited state populations are consistent with those expected for gas at a single pressure not much higher than the values ``typically'' found for local diffuse clouds --- i.e., no high-pressure component is required.
If $T$ = 100 K, then the local hydrogen densities would have been about 20 cm$^{-3}$ in 1995 and about 45--34 cm$^{-3}$ in 2001--2004 --- slightly higher than the limits estimated for several other sight lines with observed variations in $N$(\ion{Na}{1}) but much smaller than the values that would be inferred from ``typical'' \ion{Na}{1}/H ratios.

The variations in both the $N$(\ion{C}{1})/$N$(\ion{C}{2}) ratio and the \ion{C}{2} fine-structure excitation imply that the electron density $n_e$ increased by a factor of a few between 1995 and 2001, then declined somewhat by 2003--2004.
The higher values of $N$(\ion{C}{2}*)/$N$(\ion{C}{2}) found in 2001-2004 imply that the temperature in the $-$38 km~s$^{-1}$ component must be greater than about 50 K.
While the values for $n_e$ derived from the \ion{C}{2} excitation are lower by factors of 2.5--8.5 than the corresponding values estimated from $N$(\ion{C}{1})/$N$(\ion{C}{2}) under the assumption of photoionization equilibrium, they still indicate that hydrogen is partially ionized. 
The high observed abundances of the various trace species and \ion{C}{2}* (which is excited primarily via collisions with electrons) may be ascribed to the relatively high fractional ionization, and the variations in the column densities of \ion{Na}{1}, other trace species, and \ion{C}{2}* appear to be due to variations in density and/or ionization --- and not $N$(H) --- on scales of tens of AU.
It is not clear, however, how that ionization is produced --- or how the observed differences could be maintained over such small spatial/temporal scales.
In any case, the variable component at $-$38 km~s$^{-1}$ toward HD~219188 is not a very dense clump or filament (although it is much thicker than it is wide), but may be related to a population of cold, low column density clouds (perhaps due to turbulence) recently detected in \ion{H}{1} absorption.

\acknowledgements

We are grateful to D. York, J. Lauroesch, P. Sonnentrucker, and J. Barentine for obtaining most of the more recent optical spectra; to J. Thorburn for extracting the ARCES spectra; and to the organizers of and participants in the conference on Small Ionized and Neutral Structures in the Diffuse ISM (2006; Socorro, NM) for many stimulating presentations and discussions.  
Support for this work has been provided by NASA through grant HST-GO-09065.01-A (administered by STScI) to the University of Chicago.

Facilities: \facility{HST (GHRS, STIS)}, \facility{ARC}, \facility{KPNO:coud\'{e} feed}, \facility{ESO: 3.6m}


\clearpage

\begin{deluxetable}{llcc}
\tabletypesize{\scriptsize}
\tablecolumns{4}
\tablecaption{Optical Data \label{tab:datopt}}
\tablewidth{0pt}

\tablehead{
\multicolumn{1}{l}{Date}& 
\multicolumn{1}{l}{Facility}&
\multicolumn{1}{c}{FWHM\tablenotemark{a}}& 
\multicolumn{1}{c}{Obs.\tablenotemark{b}}\\ 
\multicolumn{2}{c}{ }&
\multicolumn{1}{c}{(km s$^{-1}$)}&
\multicolumn{1}{c}{ }}

\startdata
1980.75            & McDonald (2.7m)       &  5.9/5.4  & CEA \\
1991.56            & ESO (CAT/CES)         &  4.4/4.4  & KRS \\
 & \\
1995.82            & KPNO (coud\'{e} feed) &  .../1.40 & ELF \\
1997.77            & KPNO (coud\'{e} feed) & 1.35/1.35 & DEW \\
1998.68            & AAO (AAT/UCLES)       &  5.0/...  & DEW \\
1999.42            & KPNO (coud\'{e} feed) & 1.50/...  & DEW \\
1999.98            & ESO (3.6m/CES)        & 1.20/...  & DEW \\
2000.46            & KPNO (coud\'{e} feed) & 1.50/1.50 & DEW \\
 & \\
2000.80            & ESO (3.6m/CES)        &  1.3/2.0  & DEW \\
2001.86            & KPNO (coud\'{e} feed) &  4.0/...  & JTL \\
2002.79            & APO (3.5m/ARCES)      &  8.0/8.0  &  PS \\
2002.82            & APO (3.5m/ARCES)      &  8.0/8.0  &  PS \\
2002.87            & KPNO (coud\'{e} feed) &  .../4.0  & JTL \\
2003.04            & APO (3.5m/ARCES)      &  8.0/8.0  & DGY \\
2003.70            & APO (3.5m/ARCES)      &  8.0/8.0  & DGY \\
2003.74            & APO (3.5m/ARCES)      &  8.0/8.0  & DGY \\
2003.85            & APO (3.5m/ARCES)      &  8.0/8.0  &  JB \\
2003.99            & APO (3.5m/ARCES)      &  8.0/8.0  & DGY \\
2006.93            & APO (3.5m/ARCES)      &  8.0/8.0  & DGY \\
\enddata
\tablecomments{First two entries are from the literature; next six entries are from Welty \& Fitzpatrick (2001); last 11 entries are new observations.}
\tablenotetext{a}{FWHM values are for Na~I and Ca~II, respectively.}
\tablenotetext{b}{Observer: CEA = C. E. Albert; KRS = K. R. Sembach; ELF = E. L. Fitzpatrick; DEW = D. E. Welty; JTL = J. T. Lauroesch; PS = P. Sonnentrucker; DGY = D. G. York; JB = J. Barentine.}
\end{deluxetable}

\clearpage

\begin{deluxetable}{lll}
\tabletypesize{\scriptsize}
\tablecolumns{3}
\tablecaption{UV Data \label{tab:datuv}}
\tablewidth{0pt}

\tablehead{
\multicolumn{1}{l}{Date}&
\multicolumn{1}{c}{Instrument}&
\multicolumn{1}{c}{Data Sets}}
\startdata
1994.43 & GHRS (ECH-B)        & Z2FF0(208T-20ET)\\
1995.37 & GHRS (ECH-A)        & Z2FF0(105T, 108T-10BT, 10ET-10GT)\\
2001.74 & STIS (E140H, E230H) & o6e701010, o6e701020, o6e701030\\
2003.43 & STIS (E140H, E230H) & o8dp01010, o8dp01020, o8dp01030\\
2004.42 & STIS (E140H, E230H) & o8sw01010, o8sw01020, o8sw01030\\
\enddata
\tablecomments{The GHRS data were obtained under GTO program 5093 (L. Spitzer, PI); the STIS data were obtained under GO programs 9065, 9331, and 9962 (D. E. Welty, PI).}
\end{deluxetable}

\clearpage

\begin{deluxetable}{llc}
\tabletypesize{\scriptsize}
\tablecolumns{3}
\tablecaption{UV Absorption Lines\label{tab:uvlines}}
\tablewidth{0pt}

\tablehead{
\multicolumn{1}{c}{Species}&
\multicolumn{1}{c}{$\lambda$}&
\multicolumn{1}{c}{log($f\lambda$)}}
\startdata
C I    & 1188.8329 & 1.169 \\
       & 1193.0300 & 1.688 \\
       & 1193.9954 & 1.172 \\
       & 1260.7351 & 1.806 \\
       & 1276.4825 & 0.876 \\
       & 1277.2454 & 2.037 \\
       & 1280.1353 & 1.527 \\
       & 1328.8333 & 2.003 \\
       & 1560.3092 & 2.082 \\
C II   & 1334.5323 & 2.234 \\
C II*  & 1335.6627 & 1.234 \\
       & 1335.7077 & 2.188 \\
N I    & 1199.5496 & 2.199 \\
       & 1200.2233 & 2.018 \\
       & 1200.7098 & 1.715 \\
O I    & 1302.1685 & 1.796 \\
Mg I   & 2852.9631 & 3.718 \\
Mg II  & 1239.9253 &$-$0.106 \\
       & 1240.3947 &$-$0.355 \\
       & 2796.3543 & 3.236 \\
       & 2803.5315 & 2.933 \\
Si II  & 1190.4158 & 2.541 \\
       & 1193.2897 & 2.842 \\
       & 1260.4221 & 3.171 \\
       & 1304.3702 & 2.052 \\
P II   & 1301.8743 & 1.219 \\
S I    & 1295.6531 & 2.052 \\
S II   & 1250.578  & 0.832 \\
       & 1253.805  & 1.136 \\
       & 1259.518  & 1.320 \\
S III  & 1190.203  & 1.449 \\
Cl I   & 1188.7515 & 0.815 \\
       & 1188.7742 & 1.921 \\
       & 1347.2396 & 2.314 \\
Cr II  & 2062.2361 & 2.194 \\
Mn II  & 1197.184  & 2.414 \\
       & 2606.462  & 2.712 \\
Fe I   & 2484.0209 & 3.131 \\
Fe II  & 2249.8768 & 0.612 \\
       & 2260.7805 & 0.742 \\
       & 2344.2139 & 2.427 \\
       & 2374.4612 & 1.871 \\
       & 2382.7652 & 2.882 \\
       & 2600.1729 & 2.793 \\
Ni II  & 1317.217  & 2.009 \\
Zn II  & 2026.1370 & 3.007 \\
\enddata
\tablecomments{Rest wavelengths and $f$-values are from Morton (2003), except for Ni II $\lambda$1317, which has $f$ from Welty et al. (1999b).}
\end{deluxetable}

\clearpage

\begin{deluxetable}{rrcrrcrrr}
\tabletypesize{\scriptsize}
\tablecolumns{9}
\tablecaption{Component Structures\label{tab:comp}}
\tablewidth{0pt}

\tablehead{
\multicolumn{1}{c}{Component}&
\multicolumn{1}{c}{$v$(Na I)}&
\multicolumn{1}{c}{$b$(Na I)}&
\multicolumn{1}{c}{$N$(Na I)}&
\multicolumn{1}{c}{$v$(Ca II)}&
\multicolumn{1}{c}{$b$(Ca II)}&
\multicolumn{1}{c}{$N$(Ca II)}&
\multicolumn{1}{c}{$N$(Fe II)\tablenotemark{a}}&
\multicolumn{1}{c}{$N$(S II)\tablenotemark{a}}\\
\multicolumn{1}{c}{ }&
\multicolumn{1}{c}{(km~s$^{-1}$)}&
\multicolumn{1}{c}{(km~s$^{-1}$)}&
\multicolumn{1}{c}{(10$^{11}$ cm$^{-2}$)}&
\multicolumn{1}{c}{(km~s$^{-1}$)}&
\multicolumn{1}{c}{(km~s$^{-1}$)}&
\multicolumn{1}{c}{(10$^{10}$ cm$^{-2}$)}&
\multicolumn{1}{c}{(10$^{13}$ cm$^{-2}$)}&
\multicolumn{1}{c}{(10$^{14}$ cm$^{-2}$)}}

\startdata
1  & \nodata &\nodata& \nodata     & $-$51.7 & (2.5) &  \nodata     &  0.02$\pm$0.00 &  \nodata \\
2  & \nodata &\nodata& \nodata     & $-$47.0 & (2.5) &  \nodata     &  0.04$\pm$0.00 &  \nodata \\
3  & \nodata &\nodata& \nodata     & $-$41.0 & (2.5) &  \nodata     &  0.09$\pm$0.01 &  0.03$\pm$0.01 \\
4  & $-$38.3 &  0.58 & variable    & $-$38.3 & (1.0) & variable     &  0.22$\pm$0.01 &  0.07$\pm$0.01 \\
5  & \nodata &\nodata& \nodata     & $-$34.0 & (2.5) &  1.2$\pm$0.5 &  0.61$\pm$0.03 &  0.16$\pm$0.02 \\
6  & $-$30.2 & (4.0) & 0.2$\pm$0.1 & $-$29.7 & (2.5) & 12.8$\pm$0.6 &  2.58$\pm$0.56 &  0.55$\pm$0.04 \\
7  & \nodata &\nodata& \nodata     & $-$27.0 & (2.5) &  9.5$\pm$0.6 &  1.32$\pm$0.40 &  0.29$\pm$0.04 \\
8  & \nodata &\nodata& \nodata     & $-$22.5 & (2.8) & 11.0$\pm$0.5 &  3.10$\pm$0.20 &  0.55$\pm$0.02 \\
9  & \nodata &\nodata& \nodata     & $-$18.1 & (2.0) &  5.8$\pm$0.4 &  1.48$\pm$0.13 &  0.36$\pm$0.01 \\
 & \\
10 & $-$14.2 &  1.25 & 3.8$\pm$0.2 & $-$14.4 & (1.5) &  8.8$\pm$0.4 &  1.66$\pm$0.13 & (5.50) \\
11 & \nodata &\nodata& \nodata     & $-$12.1 & (1.2) &  5.4$\pm$0.5 &  2.40$\pm$0.22 & (7.70) \\
12 & $-$10.1 & (1.3) & 9.1$\pm$0.6 & $-$10.0 & (1.0) & 25.4$\pm$1.0 &  6.27$\pm$0.73 &(34.10) \\
13 &  $-$7.8 & (1.0) &16.2$\pm$2.0 &  $-$7.9 & (0.9) & 45.0$\pm$1.3 & 14.09$\pm$0.73 &(20.90) \\
14 & \nodata &\nodata& \nodata     &  $-$5.6 & (1.0) & 30.4$\pm$0.8 &  7.14$\pm$0.54 &(11.00) \\
15 &  $-$4.5 & (1.5) & 7.1$\pm$0.4 &  $-$3.6 & (1.0) & 19.2$\pm$0.7 &  7.25$\pm$0.59 &(16.50) \\
16 &  $-$1.6 &  1.25 & 7.9$\pm$0.5 &  $-$1.7 & (1.0) & 17.4$\pm$0.6 &  7.38$\pm$0.64 & (9.90) \\
17 & \nodata &\nodata& \nodata     &     0.7 & (1.5) &  7.4$\pm$0.4 &  2.55$\pm$0.22 & (4.40) \\
18 &     2.1 &  1.25 & 1.3$\pm$0.2 &     2.8 & (1.5) &  4.1$\pm$0.4 &  2.01$\pm$0.18 & (1.10) \\
 & \\
19 & \nodata &\nodata& \nodata     &     5.4 & (1.2) &  1.0$\pm$0.3 &  0.18$\pm$0.01 &  0.04$\pm$0.01 \\
20 & \nodata &\nodata& \nodata     &     8.1 & (1.5) &  \nodata     &  0.18$\pm$0.01 &  0.03$\pm$0.01 \\
21 & \nodata &\nodata& \nodata     &    11.0 & (1.5) &  \nodata     &  0.07$\pm$0.01 &  0.01$\pm$0.01 \\
22 & \nodata &\nodata& \nodata     &    13.4 & (2.5) &  \nodata     &  0.21$\pm$0.01 &  \nodata \\
23 & \nodata &\nodata& \nodata     &    19.2 & (2.5) &  \nodata     &  0.07$\pm$0.01 &  \nodata \\
24 & \nodata &\nodata& \nodata     &    43.6 & (2.0) &  \nodata     &  0.01$\pm$0.00 &  \nodata \\
\enddata
\tablecomments{The three groups of components are negative intermediate-velocity, low-velocity, and positive intermediate-velocity.  Values in parentheses were fixed in the profile fits.}
\tablenotetext{a}{Fe~II and S~II column densities derived from average STIS spectra.  Velocities of several of the IV components are slightly different for S~II.}
\end{deluxetable}

\clearpage

\begin{deluxetable}{lccccccc}
\tabletypesize{\scriptsize}
\tablecolumns{8}
\tablecaption{Variations in Na~I and Ca~II ($v$ = $-$38 km s$^{-1}$ Component) \label{tab:naca}}
\tablewidth{0pt}

\tablehead{
\multicolumn{1}{c}{ }&
\multicolumn{3}{c}{-~-~-~-~-~-~Na I~-~-~-~-~-~-}&
\multicolumn{3}{c}{-~-~-~-~-~-~Ca II~-~-~-~-~-~-}&
\multicolumn{1}{c}{ }\\
\multicolumn{1}{l}{Date}& 
\multicolumn{1}{c}{$N$}& 
\multicolumn{1}{c}{$b$}& 
\multicolumn{1}{c}{$v$}&
\multicolumn{1}{c}{$N$}&
\multicolumn{1}{c}{$b$}&
\multicolumn{1}{c}{$v$}&
\multicolumn{1}{c}{$N$(Na I)/}\\ 
\multicolumn{1}{c}{ }&
\multicolumn{1}{c}{(10$^{10}$ cm$^{-2}$)}&  
\multicolumn{1}{c}{(km s$^{-1}$)}&  
\multicolumn{1}{c}{(km s$^{-1}$)}&
\multicolumn{1}{c}{(10$^{10}$ cm$^{-2}$)}&  
\multicolumn{1}{c}{(km s$^{-1}$)}&  
\multicolumn{1}{c}{(km s$^{-1}$)}&
\multicolumn{1}{c}{$N$(Ca II)}}

\startdata
1980.75\tablenotemark{a} & $<$3.0   & \nodata       & \nodata         &    4$\pm$1   & (1.5) & ($-$38.6)       & $<$0.8      \\
1991.56\tablenotemark{b} &  4$\pm$3 &  1.5$\pm$0.9  & $-$37.9$\pm$1.1 &    2$\pm$1   & (1.5) & ($-$38.6)       & 2.0$\pm$1.8 \\
 & \\
1995.82                  & \nodata  & \nodata       & \nodata         &  2.8$\pm$0.7 & (1.0) & $-$38.3$\pm$0.1 & \nodata     \\
1997.77                  & 27$\pm$1 & 0.59$\pm$0.06 & $-$38.3$\pm$0.1 &  4.9$\pm$0.8 & (1.0) & $-$38.3$\pm$0.1 & 5.5$\pm$0.9 \\
1998.68                  & 33$\pm$2 &     (0.6)     & $-$38.0$\pm$0.1 & \nodata      &\nodata& \nodata         & \nodata     \\
1999.42                  & 40$\pm$4 &     (0.6)     & $-$38.1$\pm$0.1 & \nodata      &\nodata& \nodata         & \nodata     \\
1999.98                  & 53$\pm$2 & 0.55$\pm$0.05 & $-$38.2$\pm$0.1 & \nodata      &\nodata& \nodata         & \nodata     \\
2000.46                  & 60$\pm$3 & 0.56$\pm$0.09 & $-$38.3$\pm$0.1 & 11.6$\pm$1.2 & (1.0) & $-$38.1$\pm$0.1 & 5.2$\pm$0.6 \\
 & \\
2000.80                  & 49$\pm$1 & 0.58$\pm$0.03 & $-$38.3$\pm$0.1 &  5.3$\pm$0.7 & (1.0) & $-$38.3$\pm$0.1 & 9.2$\pm$1.2 \\
2001.86                  & 38$\pm$2 &     (0.58)    & $-$38.3$\pm$0.1 & \nodata      &\nodata& \nodata         & \nodata     \\
2002.79                  & 36$\pm$2 &     (0.58)    & $-$37.3$\pm$0.1 &  5.7$\pm$1.0 & (1.0) & $-$38.3$\pm$0.1 & 6.3$\pm$1.2 \\
2002.82                  & 33$\pm$2 &     (0.58)    & $-$37.4$\pm$0.1 &  6.5$\pm$1.3 & (1.0) & $-$38.3$\pm$0.1 & 5.1$\pm$1.1 \\
2002.87                  & \nodata  & \nodata       & \nodata         &  5.1$\pm$0.5 & (1.0) & $-$39.0$\pm$0.2 & \nodata     \\
2003.04                  & 32$\pm$2 &     (0.58)    & $-$37.5$\pm$0.1 &  5.4$\pm$1.9 & (1.0) & $-$39.3$\pm$0.1 & 5.9$\pm$2.1 \\
2003.70                  & 36$\pm$2 &     (0.58)    & $-$37.7$\pm$0.1 &  6.2$\pm$1.2 & (1.0) & $-$38.4$\pm$0.1 & 5.8$\pm$1.2 \\
2003.74                  & 33$\pm$2 &     (0.58)    & $-$39.1$\pm$0.1 &  5.1$\pm$1.0 & (1.0) & $-$39.8$\pm$0.1 & 6.5$\pm$1.3 \\
2003.85                  & 31$\pm$2 &     (0.58)    & $-$37.8$\pm$0.1 &  6.1$\pm$1.1 & (1.0) & $-$38.5$\pm$0.1 & 5.1$\pm$1.0 \\
2003.99                  & 29$\pm$2 &     (0.58)    & $-$38.6$\pm$0.1 &  6.1$\pm$1.1 & (1.0) & $-$39.7$\pm$0.1 & 4.8$\pm$0.9 \\
2006.93                  & 27$\pm$2 &     (0.58)    & $-$38.2$\pm$0.1 &  5.2$\pm$1.0 & (1.0) & $-$38.3$\pm$0.1 & 5.2$\pm$1.1 \\
\enddata
\tablecomments{First two entries are from the literature; next six entries are from Welty \& Fitzpatrick (2001); last 11 entries are new observations.
Uncertainties are 1$\sigma$; limits are 3$\sigma$; values in parentheses were fixed in the fits.}
\tablenotetext{a}{Albert (1983); $N$(Ca II) is from new fit to generated spectrum.}
\tablenotetext{b}{Sembach et al. (1993); shifted to align low-velocity absorption; $N$(Ca II) is from new fit to generated spectrum.}
\end{deluxetable}

\clearpage

\begin{deluxetable}{lrrrrrrrcr}
\tabletypesize{\scriptsize}
\tablecolumns{10}
\tablecaption{Average Column Densities and Ratios (2001.74--2004.42)\label{tab:ivlv}}
\tablewidth{0pt}

\tablehead{
\multicolumn{1}{c}{Species}&
\multicolumn{1}{c}{A$_{\odot}$\tablenotemark{a}}&
\multicolumn{1}{c}{Halo\tablenotemark{b}}&
\multicolumn{1}{c}{Warm\tablenotemark{b}}&
\multicolumn{1}{c}{Cold\tablenotemark{b}}&
\multicolumn{1}{c}{$-$38 km~s$^{-1}$\tablenotemark{c}}&
\multicolumn{1}{c}{IV\tablenotemark{c}}&
\multicolumn{1}{c}{LV\tablenotemark{c}}&
\multicolumn{1}{c}{D(LV)}&
\multicolumn{1}{c}{Total\tablenotemark{c}}}

\startdata
H (total)             &  12.00 & \nodata& \nodata& \nodata& [17.78]         & [19.21]         &  20.86$\pm$0.17 & \nodata &  20.87$\pm$0.17 \\
C I\tablenotemark{d}  &   8.39 &$-$0.20 &$-$0.20 &$-$0.20 &  13.55$\pm$0.04 &  11.79$\pm$0.23 &  13.97$\pm$0.02 & \nodata &  14.11$\pm$0.02 \\ 
N I                   &   7.83 &$-$0.05 &$-$0.05 &$-$0.05 &  12.28$\pm$0.06 &  14.01$\pm$0.01 & (16.65)         &($-$0.04)& (16.65)         \\ %
O I                   &   8.69 &$-$0.10 &$-$0.10 &$-$0.20 &  13.81$\pm$0.17 & (15.30)         &  17.30$\pm$0.08 & $-$0.25 &  17.30$\pm$0.08 \\ %
Na I                  &   6.30 & \nodata& \nodata&$-$0.60 &  11.53$\pm$0.03 &  10.30$\pm$0.24 &  12.66$\pm$0.02 & \nodata &  12.69$\pm$0.02 \\ 
Mg I                  &   7.55 &$-$0.30 &$-$0.55 &$-$1.25 &  11.97$\pm$0.08 &  11.93$\pm$0.02 & (13.10)         & \nodata & (13.16)         \\ 
Mg II                 & \nodata& \nodata& \nodata& \nodata&  $<$13.88       &  \nodata        &  15.60$\pm$0.02 & $-$0.81 &  15.60$\pm$0.02 \\ %
Si II\tablenotemark{e}&   7.54 &$-$0.30 &$-$0.40 &$-$1.30 &  12.73$\pm$0.04 &  14.43$\pm$0.08 & (15.65)         &($-$0.75)& (15.68)         \\ 
P II                  &   5.46 &$-$0.10 &$-$0.20 &$-$0.85 &  $<$12.36       &  \nodata        &  14.12$\pm$0.01 & $-$0.20 &  14.12$\pm$0.01 \\ %
S I                   &   7.19 &   0.00 &   0.00 &        &  $<$11.68       &  \nodata        &  11.71$\pm$0.21 & \nodata &  11.71$\pm$0.21 \\ %
S II                  & \nodata& \nodata& \nodata& \nodata&  12.87$\pm$0.03 &  14.28$\pm$0.01 & (16.05)         &   (0.00)& (16.06)         \\ %
S III                 & \nodata& \nodata& \nodata& \nodata&  12.15$\pm$0.33 &  13.80$\pm$0.02 & (13.76)         & \nodata & (14.09)         \\ %
Cl I                  &   5.26 & \nodata&$-$0.10 &$-$0.60 &  $<$11.48       &  \nodata        &  13.56$\pm$0.06 & \nodata &  13.56$\pm$0.06 \\ %
Ca II                 &   6.34 &$-$0.80 &$-$2.00 &$-$3.60 &  10.76$\pm$0.07 &  11.61$\pm$0.02 &  12.21$\pm$0.01 & \nodata &  12.32$\pm$0.01 \\ 
Mn II                 &   5.50 &$-$0.70 &$-$0.95 &$-$1.60 &  $<$11.58       &  11.86$\pm$0.10 &  12.96$\pm$0.02 & $-$1.40 &  13.00$\pm$0.03 \\ %
Fe I                  &   7.47 &$-$0.55 &$-$1.25 &$-$2.15 &  $<$10.60       &  \nodata        &  11.02$\pm$0.12 & \nodata &  11.02$\pm$0.12 \\ %
Fe II                 & \nodata& \nodata& \nodata& \nodata&  12.34$\pm$0.02 &  13.97$\pm$0.03 &  14.71$\pm$0.01 & $-$1.62 &  14.78$\pm$0.01 \\ %
Ni II                 &   6.22 &$-$0.60 &$-$1.40 &$-$2.25 &  $<$11.72       &  12.74$\pm$0.03 &  13.37$\pm$0.01 & $-$1.70 &  13.48$\pm$0.01 \\ %
Cu II                 &   4.26 & \nodata&$-$1.10 &$-$1.40 &  $<$11.18       &  \nodata        &  12.20$\pm$0.05 & $-$0.92 &  12.20$\pm$0.05 \\ %
Ge II                 &   3.62 & \nodata&$-$0.60 &$-$1.10 &  $<$10.56       &  \nodata        &  11.87$\pm$0.02 & $-$0.61 &  11.87$\pm$0.02 \\ %
& \\
C I/Na I              & \nodata& \nodata& \nodata& \nodata&   2.02$\pm$0.05 &   1.48$\pm$0.30 &   1.32$\pm$0.03 & \nodata &   1.43$\pm$0.02 \\ %
Mg I/Na I             & \nodata& \nodata& \nodata& \nodata&   0.44$\pm$0.08 &   1.63$\pm$0.24 &  (0.44)         & \nodata &  (0.47)         \\ %
Cl I/Na I             & \nodata& \nodata& \nodata& \nodata&$<-0.05$         &  \nodata        &   0.90$\pm$0.06 & \nodata &   0.87$\pm$0.06 \\ %
Ca II/Na I            & \nodata& \nodata& \nodata& \nodata&$-$0.77$\pm$0.07 &   1.31$\pm$0.24 &$-$0.45$\pm$0.02 & \nodata &$-$0.37$\pm$0.02 \\ %
 & \\
S II/S III            & \nodata& \nodata& \nodata& \nodata&   0.72$\pm$0.37 &   0.48$\pm$0.03 &  (2.29)         & \nodata &  (1.97)         \\ %
N I/S II              &   0.64 &   0.59 &   0.59 &   0.59 &$-$0.59$\pm$0.09 &$-$0.27$\pm$0.02 &  (0.60)         & \nodata &  (0.60)         \\ %
O I/S II              &   1.50 &   1.40 &   1.40 &   1.30 &   0.94$\pm$0.18 &  (1.02)         &  (1.25)         & \nodata &  (1.25)         \\ %
 & \\
Si II/S II            &   0.35 &   0.05 &$-$0.05 &$-$0.95 &$-$0.14$\pm$0.08 &   0.15$\pm$0.04 &($-0.40$)        & \nodata & ($-0.37$)        \\ %
Fe II/S II            &   0.28 &$-$0.27 &$-$0.97 &$-$1.87 &$-$0.53$\pm$0.07 &$-$0.31$\pm$0.04 &($-1.34$)        & \nodata & ($-1.27$)        \\ %
Ni II/S II            &$-$0.97 &$-$1.57 &$-$2.37 &$-$3.22 &$<-1.14$         &$-$1.54$\pm$0.03 &($-2.68$)        & \nodata & ($-2.57$)        \\ %
\enddata
\tablecomments{All values are log[$N$ (cm$^{-2}$)] or log(ratio).  Column densities in parentheses are approximate values estimated from saturated lines.}
\tablenotetext{a}{Solar abundances from Lodders (2003).}
\tablenotetext{b}{Values for representative ``halo'', ``warm disk'' and ``cold disk'' diffuse clouds.  First section gives depletions adopted from Welty et al. (1999b), Jenkins (2004b), and Cartledge et al. (2006).}
\tablenotetext{c}{Values are for average of STIS spectra (3 epochs).  IV does not include $-$38 km~s$^{-1}$ component.}
\tablenotetext{d}{Values are from $\lambda$1328 multiplet.}
\tablenotetext{e}{Values are from $\lambda$1304 line.}
\end{deluxetable}

\clearpage

\begin{deluxetable}{lrrrrrrl}
\tabletypesize{\scriptsize}
\tablecolumns{8}
\tablecaption{Average C I Fine-Structure Excitation ($v$ = $-$38 km s$^{-1}$ Component) \label{tab:c1fs}}
\tablewidth{0pt}

\tablehead{
\multicolumn{1}{l}{Multiplet}&
\multicolumn{1}{c}{$N$(C I)}&
\multicolumn{1}{c}{$N$(C I*)}&
\multicolumn{1}{c}{$N$(C I**)}&
\multicolumn{1}{c}{$N$(C I$_{\rm tot}$)}&
\multicolumn{1}{c}{$f_1$\tablenotemark{a}}&
\multicolumn{1}{c}{$f_2$\tablenotemark{a}}&
\multicolumn{1}{l}{Comments}\\
\multicolumn{1}{c}{ }&
\multicolumn{1}{c}{(10$^{13}$ cm$^{-2}$)}&
\multicolumn{1}{c}{(10$^{13}$ cm$^{-2}$)}&
\multicolumn{1}{c}{(10$^{13}$ cm$^{-2}$)}&
\multicolumn{1}{c}{(10$^{13}$ cm$^{-2}$)}&
\multicolumn{1}{c}{ }&
\multicolumn{1}{c}{ }&
\multicolumn{1}{c}{ }
}
\startdata
\multicolumn{8}{c}{GHRS (1995.37)} \\
\tableline
1193 & 1.12$\pm$0.19 &(0.20)         & \nodata       & \nodata       & \nodata       & \nodata       & blend with Si II \\ 
1194 & 0.86$\pm$0.33 & \nodata       & \nodata       & \nodata       & \nodata       & \nodata       & \nodata          \\ 
1560 & 0.96$\pm$0.15 & 0.24$\pm$0.03 & 0.07$\pm$0.03 & 1.27$\pm$0.16 & 0.19$\pm$0.03 & 0.06$\pm$0.02 & \nodata          \\ 
\tableline
\multicolumn{8}{c}{STIS (2001.74--2004.42)} \\
\tableline
1188 & 3.45$\pm$0.37 & \nodata       & \nodata       & \nodata       & \nodata       & \nodata       & blend with Cl I  \\ 
1193 & 2.80$\pm$0.37 & 1.28$\pm$0.15 &(0.27)         & 4.35$\pm$0.40 & 0.29$\pm$0.04 &(0.06)         & blend with Si II \\ 
1194 & 2.97$\pm$0.25 & 1.34$\pm$0.40 &(0.27)         & 4.58$\pm$0.47 & 0.29$\pm$0.09 &(0.06)         & \nodata          \\ 
1260 & \nodata       & 0.88$\pm$0.05 & 0.21$\pm$0.03 & \nodata       & \nodata       & \nodata       & blend with Si II \\ 
1276 & 2.74$\pm$0.22 & 1.78$\pm$0.34 &(0.27)         & 4.79$\pm$0.40 & 0.37$\pm$0.08 &(0.06)         & weak             \\ 
1277 & 2.59$\pm$0.42 & 0.99$\pm$0.06 & 0.22$\pm$0.02 & 3.80$\pm$0.42 & 0.26$\pm$0.03 & 0.06$\pm$0.01 & \nodata          \\ 
1280 & 2.91$\pm$0.17 & 1.24$\pm$0.07 & 0.30$\pm$0.06 & 4.45$\pm$0.19 & 0.28$\pm$0.02 & 0.07$\pm$0.01 & \nodata          \\ 
1328 & 2.40$\pm$0.35 & 0.93$\pm$0.05 & 0.20$\pm$0.02 & 3.53$\pm$0.35 & 0.26$\pm$0.03 & 0.06$\pm$0.01 & \nodata          \\ 
\enddata
\tablecomments{Values in parentheses were assumed in the fits.}
\tablenotetext{a}{$f_1$ = $N$(C I*)/$N$(C I$_{\rm tot}$); $f_2$ = $N$(C I**)/$N$(C I$_{\rm tot}$)}
\end{deluxetable}

\clearpage

\begin{deluxetable}{lcccccc}
\tabletypesize{\scriptsize}
\tablecolumns{7}
\tablecaption{Variations in Column Densities ($v$ = $-$38 km s$^{-1}$ Component) \label{tab:cduv}}
\tablewidth{0pt}

\tablehead{
\multicolumn{1}{l}{Ion}&
\multicolumn{1}{c}{GHRS B}&
\multicolumn{1}{c}{GHRS A}&
\multicolumn{1}{c}{STIS 1}&
\multicolumn{1}{c}{STIS 2}&
\multicolumn{1}{c}{STIS 3}&
\multicolumn{1}{c}{STIS avg}\\
\multicolumn{1}{c}{ }&
\multicolumn{1}{c}{1994.43}&
\multicolumn{1}{c}{1995.37}&
\multicolumn{1}{c}{2001.74}&
\multicolumn{1}{c}{2003.43}&
\multicolumn{1}{c}{2004.42}&
\multicolumn{1}{c}{ }\\
\multicolumn{1}{c}{ }&
\multicolumn{1}{c}{(cm$^{-2}$)}&
\multicolumn{1}{c}{(cm$^{-2}$)}&
\multicolumn{1}{c}{(cm$^{-2}$)}&
\multicolumn{1}{c}{(cm$^{-2}$)}&
\multicolumn{1}{c}{(cm$^{-2}$)}&
\multicolumn{1}{c}{(cm$^{-2}$)}}
\startdata
C I\tablenotemark{a}          & \nodata         & 0.96$\pm$0.15e13 & 2.46$\pm$0.43e13 & 2.11$\pm$0.33e13 & 2.02$\pm$0.29e13 & 2.14$\pm$0.33e13 \\
C I*\tablenotemark{a}         & \nodata         & 0.24$\pm$0.03e13 & 1.19$\pm$0.07e13 & 0.82$\pm$0.06e13 & 0.78$\pm$0.06e13 & 0.90$\pm$0.06e13 \\
C I**\tablenotemark{a}        & \nodata         & 0.07$\pm$0.03e13 & 0.29$\pm$0.04e13 & 0.15$\pm$0.03e13 & 0.19$\pm$0.03e13 & 0.20$\pm$0.03e13 \\
C I$_{\rm tot}$               & \nodata         & 1.27$\pm$0.16e13 & 3.94$\pm$0.44e13 & 3.08$\pm$0.34e13 & 2.99$\pm$0.30e13 & 3.24$\pm$0.34e13 \\
 & \\
C II\tablenotemark{b}         & \nodata         & (4.8$\pm$2.1e13) & (3.3$\pm$1.4e13) & (3.8$\pm$1.2e13) & (3.8$\pm$1.5e13) & (3.8$\pm$1.3e13) \\
C II*                         & \nodata         &  0.4$\pm$0.1e13  &  1.5$\pm$0.3e13  &  1.3$\pm$0.2e13  &  0.9$\pm$0.2e13  &  1.2$\pm$0.2e13  \\
C II$_{\rm tot}$\tablenotemark{b}& \nodata      & (5.2$\pm$2.1e13) & (4.8$\pm$1.3e13) & (5.1$\pm$1.2e13) & (4.7$\pm$1.5e13) & (5.0$\pm$1.3e13) \\
 & \\
N I                           & \nodata         & \nodata          &  2.0$\pm$0.3e12  &  1.8$\pm$0.3e12  &  1.7$\pm$0.3e11  &  1.8$\pm$0.3e12  \\
O I                           & \nodata         &  5.7$\pm$2.6e13  &  6.9$\pm$2.7e13  &  5.8$\pm$2.0e13  & 10.1$\pm$4.1e13  &  6.7$\pm$2.5e13  \\
Na I\tablenotemark{c}         & \nodata         & \nodata          & [3.8$\pm$0.2e11] & [3.6$\pm$0.2e11] & [2.9$\pm$0.2e11] & [3.4$\pm$0.2e11] \\
Mg I                          & \nodata         & \nodata          &  9.6$\pm$2.3e11  &  9.3$\pm$2.4e11  &  7.3$\pm$1.7e11  &  8.4$\pm$2.0e11  \\
Mg II                         & \nodata         & \nodata          & \nodata          & \nodata          & \nodata          & $<$7.5e13        \\
Si II\tablenotemark{d}        & \nodata         &  5.3$\pm$1.0e12  &  5.7$\pm$0.4e12  &  5.7$\pm$0.4e12  &  5.3$\pm$0.5e12  &  5.6$\pm$0.4e12  \\
P II                          & \nodata         & $<$3.9e12        & \nodata          & \nodata          & \nodata          & $<$2.3e12        \\
S II\tablenotemark{e}         & \nodata         &  6.2$\pm$2.0e12  &  8.4$\pm$1.2e12  &  7.9$\pm$1.1e12  &  7.4$\pm$1.4e12  &  8.0$\pm$1.2e12  \\
S III                         & \nodata         & \nodata          & \nodata          & \nodata          & \nodata          &  1.4$\pm$0.9e12  \\
Cl I                          & \nodata         & \nodata          & \nodata          & \nodata          & \nodata          & $<$3.0e11        \\
Ca II\tablenotemark{c}        & \nodata         & [2.8$\pm$0.7e10] & \nodata          & [6.2$\pm$1.2e10] & [6.1$\pm$1.1e10] & [5.8$\pm$0.9e10] \\
Cr II                         & $<$5.3e11       & \nodata          & \nodata          & \nodata          & \nodata          & \nodata          \\
Mn II                         & $<$1.3e11       & \nodata          & \nodata          & \nodata          & \nodata          & $<$3.8e11        \\
Fe II                         &  2.1$\pm$0.2e12 & \nodata          &  2.2$\pm$0.2e12  &  2.3$\pm$0.2e12  &  2.3$\pm$0.2e12  &  2.3$\pm$0.2e12  \\
Ni II                         & \nodata         & \nodata          & \nodata          & \nodata          & \nodata          & $<$5.3e11        \\
Zn II                         & $<$7.5e10       & \nodata          & \nodata          & \nodata          & \nodata          & \nodata          \\
\enddata
\tablecomments{Uncertainties are 1$\sigma$; limits are 3$\sigma$.}
\tablenotetext{a}{Values for GHRS are from $\lambda$1560 multiplet; values for STIS are from $\lambda$1328 multiplet.}
\tablenotetext{b}{Values in parentheses are estimated assuming typical interstellar C/S = 10.4 (for C I$_{\rm tot}$ + C II$_{\rm tot}$ vs. S II).}
\tablenotetext{c}{Values in square brackets for Na~I and Ca~II are for closest epoch to UV observations (but within 6 months).}
\tablenotetext{d}{Values are from $\lambda$1304 line.}
\tablenotetext{e}{GHRS value is from $\lambda$1253 only; STIS 3 value is from $\lambda$1259 and $\lambda$1250; STIS 1,2 values are from all three lines.}

\end{deluxetable}

\clearpage

\begin{deluxetable}{lccccc}
\tabletypesize{\scriptsize}
\tablecolumns{6}
\tablecaption{Variations in Ratios ($v$ = $-$38 km s$^{-1}$ Component) \label{tab:rat}}
\tablewidth{0pt}

\tablehead{
\multicolumn{1}{l}{Ratio}&
\multicolumn{1}{c}{GHRS}&
\multicolumn{1}{c}{STIS 1}&
\multicolumn{1}{c}{STIS 2}&
\multicolumn{1}{c}{STIS 3}&
\multicolumn{1}{c}{STIS avg}\\
\multicolumn{1}{c}{ }&
\multicolumn{1}{c}{1995.37}&
\multicolumn{1}{c}{2001.74}&
\multicolumn{1}{c}{2003.43}&
\multicolumn{1}{c}{2004.42}&
\multicolumn{1}{c}{ }}
\startdata
C I*/C I$_{\rm tot}$       &  0.19$\pm$0.03 &  0.30$\pm$0.04 &  0.27$\pm$0.04 &  0.26$\pm$0.03 &  0.28$\pm$0.04 \\ 
C I**/C I$_{\rm tot}$      &  0.06$\pm$0.02 &  0.07$\pm$0.01 &  0.05$\pm$0.01 &  0.06$\pm$0.01 &  0.06$\pm$0.01 \\ 
 & \\
C II*/C II\tablenotemark{a}& (0.08$\pm$0.04)& (0.45$\pm$0.21)& (0.34$\pm$0.12)& (0.24$\pm$0.11)& (0.30$\pm$0.12)\\ %
 & \\
N I/S II                   & \nodata        &  0.24$\pm$0.05 &  0.23$\pm$0.05 &  0.23$\pm$0.06 &  0.23$\pm$0.05 \\ %
O I/S II                   &   9.2$\pm$5.1  &   8.2$\pm$3.4  &   7.3$\pm$2.7  &  13.7$\pm$6.1  &   8.5$\pm$3.4  \\ %
 & \\
Na I/Ca II\tablenotemark{b}& \nodata        & \nodata        &  [5.8$\pm$1.2] &  [4.8$\pm$0.9] &   5.9$\pm$1.0  \\ 
Si II/S II                 &  0.85$\pm$0.32 &  0.68$\pm$0.11 &  0.72$\pm$0.11 &  0.72$\pm$0.15 &  0.71$\pm$0.12 \\ %
Fe II/S II                 & [0.34$\pm$0.11]&  0.26$\pm$0.04 &  0.29$\pm$0.05 &  0.31$\pm$0.06 &  0.29$\pm$0.05 \\ %
 & \\
~[O/S]                     & $-0.54\pm0.27$ & $-0.59\pm0.19$ & $-0.63\pm0.17$ & $-0.36\pm0.21$ & $-0.57\pm0.18$ \\ %
~[Si/S]                    & $-0.42\pm0.17$ & $-0.52\pm0.07$ & $-0.49\pm0.07$ & $-0.49\pm0.09$ & $-0.50\pm0.07$ \\ %
~[Fe/S]                    &[$-0.75\pm0.15$]& $-0.86\pm0.07$ & $-0.82\pm0.07$ & $-0.79\pm0.09$ & $-0.82\pm0.08$ \\ %
\enddata
\tablecomments{Uncertainties are 1$\sigma$.}
\tablenotetext{a}{Values in parentheses use $N$(C II) estimated by assuming typical interstellar C/S = 10.4 (for C I$_{\rm tot}$ + C II$_{\rm tot}$ vs. S II).}
\tablenotetext{b}{Values in square brackets are for closest epoch to UV observations (but within 6 months).}

\end{deluxetable}

\clearpage

\begin{deluxetable}{lccccc}
\tabletypesize{\scriptsize}
\tablecolumns{6}
\tablecaption{Variations in Inferred Properties ($v$ = $-$38 km s$^{-1}$ Component) \label{tab:phys}}
\tablewidth{0pt}

\tablehead{
\multicolumn{1}{l}{Quantity}&
\multicolumn{1}{c}{GHRS}&
\multicolumn{1}{c}{STIS1}&
\multicolumn{1}{c}{STIS2}&
\multicolumn{1}{c}{STIS3}&
\multicolumn{1}{c}{STIS avg}\\
\multicolumn{1}{c}{ }&
\multicolumn{1}{c}{1995.37}&
\multicolumn{1}{c}{2001.74}&
\multicolumn{1}{c}{2003.43}&
\multicolumn{1}{c}{2004.42}&
\multicolumn{1}{c}{ }}
\startdata
$N$(H)\tablenotemark{a} (cm$^{-2}$)             &   4.9$\pm$1.4e17 &   6.3$\pm$1.0e17 &   6.0$\pm$0.9e17 &   5.7$\pm$1.1e17 &   6.0$\pm$1.0e17 \\
log($n_{\rm H}T$) (cm$^{-3}$K)\tablenotemark{b} &  3.30$\pm$0.10   &  3.65$\pm$0.11   &  3.53$\pm$0.11   &  3.53$\pm$0.09   &  3.58$\pm$0.09   \\
$n_{\rm H}$ (cm$^{-3}$)\tablenotemark{b}        &    20$\pm$ 5     &    45$\pm$11     &    34$\pm$ 9     &    34$\pm$ 7     &    38$\pm$ 8     \\
thickness (AU)                                  &  1640$\pm$610    &   940$\pm$280    &  1180$\pm$350    &  1120$\pm$320    &  1060$\pm$280    \\
$n_e$(C I)\tablenotemark{c} (cm$^{-3}$)         &   5.9$\pm$2.5    &  19.6$\pm$5.8    &  14.4$\pm$3.7    &  15.2$\pm$5.0    &  15.5$\pm$4.3    \\
$n_e$(C II)\tablenotemark{d} (cm$^{-3}$)        &   0.7$\pm$0.4    &   7.6$\pm$5.0    &   4.3$\pm$2.9    &   2.5$\pm$1.8    &   3.5$\pm$2.4    \\
$n_e$/$n_{\rm H}$\tablenotemark{e}              & 0.035$\pm$0.022  &  0.17$\pm$0.12   &  0.13$\pm$0.09   & 0.074$\pm$0.055  & 0.092$\pm$0.066  \\
\enddata
\tablenotetext{a}{From $N$(S~II)+$N$(S~III).}
\tablenotetext{b}{From C I fine-structure excitation; assuming $T$ = 100 K, $n_e$/$n_{\rm H}$ = 0.1, and WJ1 radiation field.}
\tablenotetext{c}{Assuming photoionization equilibrium for carbon; $T$ = 100 K; WJ1 radiation field.}
\tablenotetext{d}{From C II fine-structure excitation equilibrium; assuming $T$ = 100 K.}
\tablenotetext{e}{Using $n_e$ from analysis of C II fine-structure.}
\end{deluxetable}

\clearpage
 
\clearpage
\thispagestyle{empty}
\setlength{\voffset}{-20mm}
 
\clearpage
\setlength{\voffset}{0mm}

\clearpage
 
\clearpage
 
\clearpage
 
\end{document}